\documentclass[aps,prc,twocolumn,10pt,superscriptaddress]{revtex4-2}

\usepackage[utf8]{inputenc}
\usepackage{hyperref}
\usepackage{amsmath,amsfonts,amssymb}
\usepackage{graphicx}
\usepackage[dvipsnames]{xcolor}

\usepackage[normalem]{ulem}  
\renewcommand\sout{\bgroup \color{black} \ULdepth=-.5ex \ULset}
\usepackage{soul}

\begin{document}

\preprint{APS/123-QED}

\title{Constraints on Kinetic Mixing of Dark Photons from Dilepton Spectra}

\author{A. W. Romero Jorge}
\email{jorge@itp.uni-frankfurt.de}
\affiliation{Frankfurt Institute for Advanced Studies (FIAS), Ruth Moufang Str. 1, 60438 Frankfurt, Germany}
 \affiliation{Institut f\"ur Theoretische Physik, Johann Wolfgang Goethe	University, Max-von-Laue-Str. 1, 60438 Frankfurt, Germany}
 \affiliation{Helmholtz Research Academy Hessen for FAIR (HFHF), GSI Helmholtz	Center for Heavy Ion Physics. Campus Frankfurt, 60438 Frankfurt, Germany.}

\author{E. Bratkovskaya}%
 \email{E.Bratkovskaya@gsi.de}
 
\affiliation{GSI Helmholtzzentrum f\"ur Schwerionenforschung GmbH, Planckstraße 1, 64291 Darmstadt, Germany}
 
\affiliation{Institut f\"ur Theoretische Physik, Johann Wolfgang Goethe	University, Max-von-Laue-Str. 1, 60438 Frankfurt, Germany}
\affiliation{Helmholtz Research Academy Hessen for FAIR (HFHF), GSI Helmholtz	Center for Heavy Ion Physics. Campus Frankfurt, 60438 Frankfurt, Germany.}

\author{T. Song} 
\affiliation{GSI Helmholtzzentrum f\"ur Schwerionenforschung GmbH, Planckstraße 1, 64291 Darmstadt, Germany}

\author{L. Sagunski} 
\affiliation{Institut f\"ur Theoretische Physik, Johann Wolfgang Goethe	University, Max-von-Laue-Str. 1, 60438 Frankfurt, Germany}

\date{\today}

\begin{abstract}

Dark photons, the hypothetical gauge bosons associated with an additional $U(1)^{\prime}$ symmetry, can couple to Standard Model particles through a small kinetic mixing parameter $\varepsilon$ with the ordinary photon. This mechanism provides a portal between the dark sector and visible matter. In this study, we present a procedure to derive theoretical upper bounds on the kinetic mixing parameter $\varepsilon^2(M_U)$ by analyzing dilepton spectra from heavy-ion collisions across a broad energy range, from SIS to LHC energies. Our analysis is based on the microscopic Parton-Hadron-String Dynamics (PHSD) transport approach, which successfully reproduces the measured dilepton spectra in $p+p$, $p+A$, and $A+A$ collisions across the same energy range. Besides the dilepton channels resulting from interactions and decays of Standard Model particles (such as mesons and baryons), the PHSD has been extended to include the decay of hypothetical dark photons into dileptons, $U \to e^+ e^-$. The production of these dark photons occurs via Dalitz decays of $\pi^0$, $\eta$, $\omega$, $\eta^{\prime}$, and $\Delta$ resonances; direct decays of $\rho$, $\omega$, and $\phi$; the kaon mode $K^+ \to \pi^+ U$; and thermal $q\bar q$ annihilation in the quark-gluon plasma. Our results show that high-precision measurements of dilepton spectra in heavy-ion collisions provide a sensitive and competitive probe of dark photons in the MeV to multi-GeV mass range. Furthermore, we quantify the experimental accuracy required to constrain the remaining viable parameter space of kinetic mixing in dark photon scenarios.
\end{abstract}

\maketitle

\section{Introduction}\label{sec:intro}

Dark matter (DM) accounts for roughly one quarter of the total energy density of the Universe, meaning that when all forms of energy: ordinary matter, radiation, dark energy are summed, about 25 \% is contributed by a non‐luminous component that interacts primarily via gravity \cite{Planck:2018vyg}.  Its existence is inferred from phenomena observed over scales ranging from kiloparsecs (galactic rotation curves) to gigaparsecs (cosmic microwave background anisotropies), a span of nearly twenty orders of magnitude \cite{Feng:2010gw}.

On galactic scales, stellar kinematics from the Gaia mission and other collaborations reveal that the Milky Way rotation curve remains flat well beyond the visible disc, confirming the need for an extended unseen dark matter halo \cite{deSalas:2019pee}.
In cluster systems, gravitational lensing in the Bullet Cluster and Abell 520 strongly suggests that the dominant mass component behaves largely collisionlessly and is non‐luminous, separating spatially from the X-ray emitting gas \cite{Clowe:2006eq,Mahdavi:2007yp}.
Additionally, models of self-interacting dark matter  have been proposed to address small-scale structure issues, such as the observed cores of dwarf galaxies and the “too-big-to-fail” problem, providing heat conduction within dark matter halos that alleviates discrepancies between cosmological simulations and observations \cite{Tulin:2017ara}.

At cosmological distances, the Planck measurement of the cosmic microwave background (CMB) angular power spectrum has constrained the non-baryonic relic density to $
\Omega_{\mathrm{DM}}h^{2}\simeq0.12\,,$
a value also required to reproduce baryon acoustic oscillation scales, galaxy two-point correlations, and cluster mass functions in large-scale $N$-body simulations \cite{Planck:2018vyg}. The global 21 cm absorption feature detected by EDGES may hint at DM baryon interactions during cosmic dawn \cite{Berlin:2018sjs}, while indirect searches such as the Galactic Center gamma-ray excess measured by Fermi-LAT \cite{Daylan:2014rsa} and the antiproton spectrum from AMS-02 \cite{Cholis:2019ejx} continue to probe DM annihilation or decay channels.

Direct‐detection experiments have pushed nuclear recoil limits below cross sections of $10^{-48}\,\mathrm{cm}^2$ for a Weakly
Interacting Massive Particle as  one of the most
promising DM candidates with mass near 30 GeV, with LUX \cite{LUX:2016ggv}, PandaX \cite{PandaX-II:2017hlx}, XENON1T \cite{XENON:2018voc}, and LZ \cite{LZ:2024zvo} setting the most stringent bounds. Intriguingly, XENON1T’s observation of an electronic-recoil excess has been interpreted as possible solar or Galactic dark photons \cite{XENON:2020rca}.

Together, these diverse lines of evidence from stellar dynamics and gravitational lensing to CMB anisotropies,
cosmological constraints,
indirect and direct searches, and heavy-ion collision data demand a new, electrically neutral, weakly interacting particle, yet its true nature remains one of the foremost mysteries in modern physics (cf. \cite{dEnterria:2022sut}).

The simplest way to couple a hidden
sector to the Standard Model (SM) is through a set of
dimension-four operators known as
portals,\cite{Alexander:2016aln,Battaglieri:2017aum,Agrawal:2021dbo}.
Portals are usually categorized by the spin of the mediator and the dimension of the interaction operator. The most widely explored, well-motivated scenarios feature renormalizable couplings involving four distinct mediator types: a spin-1 vector, a spin-½ neutrino, a scalar Higgs, and a pseudo-scalar axion.

The vector portal is arguably the easiest: a kinetic
mixing of strength $\varepsilon$ between the hypercharge field strength
$B_{\mu\nu}$ and the  field strength $F'_{\mu\nu}$ of a new abelian
$U(1)^{\prime}$ gauge boson introduces the interaction
$  \mathcal L \;\supset\; \frac{\varepsilon}{2}\,
    B_{\mu\nu}F'^{\mu\nu}\,,$ originally proposed by Holdom\,\cite{Holdom:1985ag}.  The associated gauge boson variously called dark photon, hidden photon, $U$‐boson, or $A'$ may acquire a mass $M_{U}$ via a Stueckelberg term or a dark Higgs mechanism.
The mixing angle $\varepsilon^{2}$ fixes its coupling to SM charges and
hence its decay rate to $e^{+}e^{-}$, $\mu^{+}\mu^{-}$ or hadrons
\cite{Fayet:1980ad,Fayet:2004bw,Boehm:2003hm,Pospelov:2007mp,Batell:2009di,Batell:2009yf}.
Because $\varepsilon$ can naturally lie anywhere between $10^{-2}$ and
$10^{-7}$, dark photons remain largely unconstrained by precision
electroweak tests but are still accessible in dedicated
experiments.

A striking phenomenological advantage of the vector portal is that a dark
photon can be radiated off any electromagnetic current.  Low-mass QCD
states therefore provide abundant production mechanisms: Dalitz decays of
$\pi^{0}$, $\eta$, $\eta'$, virtual‐photon emission from the $\Delta(1232)$
resonance, and direct conversion of vector mesons
$\rho$, $\omega$, $\phi$.  Experiments look for a narrow peak in the
dilepton mass spectrum above the smooth SM background
\cite{Alexander:2016aln,Battaglieri:2017aum,Beacham:2019nyx,Billard:2021uyg}.
Some unexpected signals such as the excess of electron recoils observed in XENON1T could be explained by dark photons originating from the sun or our galaxy \cite{XENON:2020rca}.

Over the past decade a broad experimental campaign has compressed the
allowed region in the $(M_{U},\varepsilon^{2})$ plane.  Beam‐dump and
fixed-target experiments, flavor factories, hadron and lepton colliders
 presently exclude
$\varepsilon^{2}\!\gtrsim\!10^{-6}$ for $M_{U}$ between
$\approx 20$ MeV and a few GeV
\cite{Fabbrichesi:2020wbt}.  The strongest limits arise from the
MAMI A1 formfactor measurement\,\cite{Merkel:2014avp}, NA48/2 kaon
decays\,\cite{Batley:2015lha}, the APEX \cite{APEX:2011dww} and HPS fixed-target programmes\,\cite{HPS:2018xkw}, $e^{+}e^{-}$ results from
BaBar\,\cite{BaBar:2009lbr,BaBar:2014zli} and KLOE
\cite{KLOE-2:2014qxg,KLOE-2:2012lii,KLOE-2:2018kqf}, as well as high-mass dimuons at LHCb \cite{LHCb:2017trq,Aaij:2019bvg}  and
CMS\,\cite{CMS:2023hwl}.  Nevertheless, sizeable gaps remain especially
near the $\rho$, $\omega$ and $\phi$ poles where laboratory bounds,
astrophysical constraints and beam-dump searches lose sensitivity.

Heavy-ion collisions provide a complementary probe by the indirect detection:
they produce large numbers of hadrons and, at collider energies, form a
thermal quark–gluon plasma (QGP) whose both quark–antiquark annihilation and hadrons can
radiate dark photons, which subsequently decays into dileptons.

Our previous studies exploited the Dalitz decays of 
$\pi^{0}$, $\eta$ and the $\Delta$ resonance to derive limits on
$\varepsilon^{2}(M_{U})$ up to $M_{U}=0.6$ GeV/$c^{2}$
\cite{Schmidt:2021hhs,Bratkovskaya:2022cch, RomeroJorge:2024eky,Jorge:2024ris}.  In the present work we extend
the analysis in three directions:

\begin{itemize}
\item[(i)] \textbf{Extended mass range:} the accessible range is extended to
      $M_{U}=2$ GeV/$c^{2}$, covering the entire light vector meson sector
      and the QGP dilepton emission.
\item[(ii)] \textbf{Additional hadronic channels:} we include the
      Dalitz transition $\omega\!\to\pi^{0}U$, direct conversions
      $\rho,\omega,\phi\!\to U$, and the kaon decay
      $K^{+}\!\to\pi^{+}U$, all of which become relevant once
      $M_{U}\!>\!m_{\eta}$.
\item[(iii)] \textbf{Partonic radiation:} at RHIC and LHC energies thermal
      $q\bar q\!\to U$ annihilation in the QGP is evaluated consistently
      with an off-shell parton spectral function.
\end{itemize}

The paper is structured as follows.  In Sec.~\ref{sec:phsd}, we outline the Parton-Hadron-String Dynamics (PHSD) transport framework, detailing its off‐shell transport equations and the treatment of both hadronic and partonic phases.  We  summarize all Standard Model dilepton production mechanisms included in PHSD, ranging from low-mass Dalitz and vector-meson decays to intermediate- and high-mass partonic channels. 
 In Sec.~\ref{sec:DPmodel} we introduce the dark photon ($U$) model, its kinetic‐mixing Lagrangian, and the formulas for its partial widths and branching ratios into leptons and hadrons.  Sec.~\ref{sec:DPproduction} describes how dark photons are produced in PHSD   through Dalitz and direct decays of light mesons and resonances, kaon and 
 quark-antiquark ($q\bar q$) annihilation    and how their subsequent $U\to e^+e^-$ decays are handled.  The procedure for obtaining theoretical constraints on the kinetic mixing parameter $\varepsilon^2(M_U)$ by comparing the summed SM+U dilepton spectra to experimental yields is detailed in Sec.~\ref{sec:results}, where we present our results for SIS, RHIC and LHC systems.  Finally, Sec.~\ref{sec:summary} offers a concise summary and outlook.

\section{Standard matter production in the PHSD transport model}\label{sec:phsd}

\subsection{PHSD approach}

The Parton-Hadron-String Dynamics \cite{Cassing:2008sv, Cassing:2008nn, Cassing:2009vt, Bratkovskaya:2011wp, Linnyk:2015rco, Moreau:2019vhw} is a microscopic, non-equilibrium transport approach that tracks both hadronic and partonic degrees of freedom throughout the space–time evolution of a relativistic heavy-ion reaction.  Starting from the first, off-shell nucleon–nucleon ($NN$) encounters, PHSD follows the formation and expansion of the quark–gluon plasma, its interactions at partonic level, the cross-over to confined matter, and the subsequent rescattering and decays of hadrons \cite{Cassing:2008sv,Cassing:2008nn,Cassing:2009vt,Bratkovskaya:2011wp}.  The dynamics are obtained by solving the Cassing-Juchem off-shell transport equations, which arise from a first order gradient expansion of the Kadanoff–Baym equations in test  particle representation \cite{Cassing:1999wx,Juchem:2004cs}.  
This formulation treats spectral functions explicitly and is therefore well suited for systems far from equilibrium \cite{Cassing:2021fkc}.

The hadronic sector builds on the earlier Hadron-String Dynamics (HSD) model \cite{Ehehalt:1996uq,Cassing:1999es}.  It contains the baryon octet and decuplet, the pseudoscalar and vector meson nonets, and higher resonances.  Multi-particle production above the baryon-baryon threshold $\sqrt{s_{BB}^{\mathrm{th}}}=2.65~\text{GeV}$, meson-baryon threshold $\sqrt{s_{mB}^{\mathrm{th}}}=2.35~\text{GeV}$ and meson-meson threshold $\sqrt{s_{mm}^{\mathrm{th}}}=1.3~\text{GeV} $ 
are described with a Lund-type string picture implemented in {\sc fritiof} 7.02 and {\sc pythia} 6.4, tuned for intermediate energies and for in-medium masses and widths \cite{Nilsson-Almqvist:1986ast,Andersson:1992iq,Sjostrand:2006za,Kireyeu:2020wou}.  In hot and dense matter, the string fragmentation function is modified to account for chiral symmetry restoration via the Schwinger mechanism, the Cronin $k_{T}$ broadening, and spectral functions that depend on momentum, temperature, and density %
\cite{Palmese:2016rtq,Song:2020clw,Bratkovskaya:2007jk}.

Strings break into leading hadrons   containing the original valence quarks   and pre-hadrons, which carry newly produced quark–antiquark pairs.  
If the local rest  frame energy density stays below the critical value $\varepsilon_{C}\simeq0.4$ GeV/fm$^{3}$ as extracted from lattice QCD \cite{Borsanyi:2022qlh}, pre-hadrons materialise into on-shell hadrons after the formation time $t_{F}=\tau_{F}\gamma$ with $\tau_{F}=0.8$ fm/$c$ and Lorentz factor $\gamma =1/\sqrt{1- \vec v^2}$, where $v$ is the velocity of a hadron $\vec v= \vec p/E$ with 3-momentum $\vec p$ and energy $E$ in the calculational frame which is chosen as the center  of  mass frame of $NN$ collisions with initial beam energy.

Above $\varepsilon_{C}$ the pre-hadrons from the color strings “melt’’ into
partons, i.e. strongly interacting off-shell quarks, antiquarks, and gluons whose properties are provided by the Dynamical Quasi-Particle Model (DQPM) \cite{Cassing:2007yg,Moreau:2019vhw,Soloveva:2019xph}. 
In the DQPM each parton carries a complex self-energy
so the real part behaves as an effective mass (squared) while the imaginary part encodes the interaction rate.  The temperature   and baryon  chemical  potential dependence of the couplings is fixed by matching the entropy density to lattice  QCD results at $\mu_{B}=0$ \cite{Aoki:2009sc,Cheng:2007jq}.  With these inputs the DQPM reproduces the lattice equation of state and provides transport coefficients $\eta/s$, $\zeta/s$, $\sigma_{0}/T$, and the baryon diffusion constant consistent with lattice extractions over the full $(T,\mu_{B})$ plane 
\cite{Fotakis:2021diq,Soloveva:2021quj}.

Elastic scatterings $qq\leftrightarrow qq$, $\bar q\bar q\leftrightarrow\bar q\bar q$, $gg\leftrightarrow gg$ and inelastic $gg\leftrightarrow g$, $q\bar q\leftrightarrow g$ are evaluated with dressed propagators and a running coupling determined inside the DQPM, ensuring detailed balance \cite{Moreau:2019vhw}.  As the fireball expands and cools, parton spectral functions sharpen and hadronization proceeds continuously near the crossover boundary.  The resulting hadronic system is then propagated with off-shell HSD dynamics, including optional self-energies for the hadrons \cite{Cassing:2003vz,Song:2020clw}.

PHSD has been benchmarked from SIS to RHIC and LHC energies, reproducing a wide array of hadronic, photon, and dilepton observables \cite{Linnyk:2015rco,Song:2018xca,Moreau:2021clr}.  The present work employs PHSD~v6.1, which combines the $(T,\mu_{B})$ dependent DQPM equation of state of PHSD \cite{Moreau:2021clr} with updated treatments of strangeness \cite{Song:2020clw}, open heavy flavor \cite{Song:2018xca}
and dilepton radiation \cite{Jorge:2025wwp}.

\subsection{Dilepton production from Standard-Model sources in PHSD }
\label{sec:dileptonSM}

Throughout the full space–time evolution of a nucleus–nucleus collision, virtual photons are emitted at every stage and convert into dilepton pairs 
\cite{Rapp:2013nxa,Linnyk:2015rco}.  The PHSD approach explicitly tracks this emission from the initial nucleon–nucleon impact, through the formation and hadronization of the quark–gluon plasma, to the late, dilute hadronic gas, thus incorporating all established dilepton sources 
\cite{Linnyk:2015rco,Jorge:2025wwp}.  

For low invariant masses, 
the dilepton yield is dominated by hadronic processes:\\

(i) Dalitz decays of
pseudoscalar mesons and baryon resonances
\,$(\pi^{0},\eta,\eta'\!\rightarrow\gamma e^{+}e^{-};
\;\omega\!\rightarrow\pi^{0}e^{+}e^{-};
\;a_{1}\!\rightarrow\pi e^{+}e^{-};
\;\Delta\!\rightarrow N e^{+}e^{-})$,\\

(ii) direct two-body decays of the light vector mesons
$\rho^{0},\omega,\phi\!\rightarrow e^{+}e^{-}$, and \\

(iii) nucleon–nucleon as well as pion–nucleon bremsstrahlung
$NN\!\rightarrow NN e^{+}e^{-}$,
$\pi N\!\rightarrow\pi N e^{+}e^{-}$.\\

Once the QGP is created, 
the partonic channels are calculated  with off–shell matrix elements evaluated using DQPM propagators and a running coupling that depends on local temperature and  baryon density.  These include annihilation and Compton–like processes,
\begin{align}
q\bar q &\;\rightarrow\;\gamma^{\ast}\;\rightarrow\;e^{+}e^{-},\\
q\bar q &\;\rightarrow\;g\,\gamma^{\ast}\;\rightarrow\;g\,e^{+}e^{-},\\
qg &\;\rightarrow\;q\,\gamma^{\ast}\;\rightarrow\;q\,e^{+}e^{-},\quad
\bar qg\;\rightarrow\;\bar q\,\gamma^{\ast}\;\rightarrow\;\bar q\,e^{+}e^{-}.
\end{align}

In the intermediate–mass region (1  $<M_{ee}<$ 3 GeV/$c^{2}$), correlated semi‐leptonic decays of open‐charm and open‐beauty mesons dominate the spectrum, with heavy‐quark transport treated as in \cite{Song:2018xca}.  

The production of dileptons from an intermediate hadronic resonance \(R\) follows a
sequence with each step implemented in PHSD using the corresponding matrix elements and off–shell spectral functions:

\begin{align}
& BB \to R\,X, \qquad mB \to R\,X, \qquad mm \to R\,X ,                                        \label{eq:prodR} \\
& R \to e^{+}e^{-}, \qquad R \to e^{+}e^{-}\,X  ,                                              \label{eq:Rdirect}\\
& R \to m\,X,\qquad\; m \to e^{+}e^{-}\,X  ,                                                   \label{eq:DalitzCascade}\\
& R \to R'\,X,\qquad R' \to e^{+}e^{-}\,X.                                                    \label{eq:RprimeCascade}
\end{align}

\noindent
Eq. (\ref{eq:prodR}) creates the resonance in baryon–baryon,
meson–baryon, or meson–meson collisions. Eqs.\,(\ref{eq:Rdirect}-\ref{eq:RprimeCascade}) describe, respectively, direct electromagnetic
conversion, Dalitz conversion via an intermediate meson, and a two-step
cascade through a secondary resonance $R'$.

Finally, PHSD employs a continuous emission or “shining” method to accumulate the dilepton yield from all resonances along their entire trajectories. 
The dilepton yield from the  resonances $R$ with electromagnetic width
$\Gamma_{e^{+}e^{-}}(M)$ is calculated by integrating over their lifetime:
\begin{equation}
\label{eq:shining}
\frac{{\rm d}N_{e^{+}e^{-}}}{{\rm d}M}
=\sum_{j=1}^{N_{\Delta M}}
 \int_{t^{i}_{j}}^{t^{f}_{j}}
 \frac{{\rm d}t}{\gamma}\,
 \frac{\Gamma_{e^{+}e^{-}}(M)}{\Delta M},
\end{equation}
where $t^{i}_{j}$ and $t^{f}_{j}$ are, respectively, the production and
absorption (or strong‐decay) times of the $j^{\text{th}}$ resonance  inside
the mass bin $\Delta M$, summered over  the number of all resonances $N_{\Delta M}$ in this bin $\Delta M$  \cite{Heinz:1991fn,Bratkovskaya:2007jk,
Bratkovskaya:2013vx,Galatyuk:2015pkq,Staudenmaier:2017vtq,Larionov:2020fnu}.
Equation\,(\ref{eq:shining}) retains the finite lifetime, the instantaneous
off-shell mass, and the local medium properties of every emitter.

\section{Modeling of the Dark Photon production }\label{sec:DPmodel}

Several  portals have been proposed to connect the dark sector with the SM: the Higgs portal (scalar mediator), the neutrino portal (fermionic mediator), the axion portal (pseudoscalar mediator), and the vector portal (dark photon) \cite{Batell:2009di}. The first three provide complementary scenarios but typically involve additional fields, couplings, and model-dependent assumptions. In this work we focus on the  dark photons, corresponding to the vector portal, which extends the SM by a single $U(1)^{\prime}$ gauge boson. The dark photons (also referred to as the $U$-bosons, $A'$, or $\phi$) are kinetically mixed with the SM photons \cite{Holdom:1985ag,Batell:2009di,Pospelov:2008zw,Essig:2013lka}, introducing only two free parameters, the mass $M_U$ and the kinetic mixing strength. This simple construction induces a small coupling to the electromagnetic current, making it possible to be potentially observed via dilepton measurements in heavy-ion collision experiments from SIS to LHC energies \cite{Alexander:2016aln}.

The general Lagrangian describing the minimal dark photon model, incorporating kinetic mixing with the SM photon, is given by:
\begin{eqnarray}
    \mathcal{L}_U =  
    -\frac{1}{4}F'_{\mu\nu}F'^{\mu\nu} + \frac{\varepsilon}{2}B_{\mu\nu}F'^{\mu\nu} - \frac{1}{2} M_U^2 A'_{\mu} A'^{\mu}, 
\end{eqnarray}
where $B_{\mu\nu}$ and $F'_{\mu\nu}$ are the field strength tensors of the SM field and the dark photon respectively, $\varepsilon$ is the kinetic mixing parameter, controlling the interaction strength between the dark photon and the SM photon and $M_U$ is the mass of the dark photon, which can arise via spontaneous symmetry breaking or be introduced explicitly.

The kinetic mixing term induces an effective interaction between the dark photon and electrically charged SM particles. In the small mixing limit ($\varepsilon \ll 1$), the dark photon behaves similarly to a SM photon but with a suppressed coupling $e\varepsilon$ to the electromagnetic current and effetive fine structure constant $\alpha^{\prime}=\varepsilon^2 \alpha $.

The dark photon can decay into SM particles if kinematically allowed, with the dominant decay channel depending on its mass. For $M_U > 2m_e$, the leading decay mode is into an electron-positron pair ($U \to e^+ e^-$) while for $M_U > 2m_\mu$ decays also in muon pairs ($U \to \mu^+ \mu^-$), both with a decay width given by \cite{Pospelov:2008zw}:
\begin{equation}
    \Gamma(U \to l^+ l^-) = \frac{1}{3} \alpha \varepsilon^2 M_U \left(1 + \frac{2m_l^2}{M_U^2}\right) \sqrt{1 - \frac{4m_l^2}{M_U^2}}, \label{BrUll}
\end{equation}
with $l=e,\mu$.
If \(M_U\) is larger, it can also decay into heavier lepton pairs or hadronic final states \cite{Essig:2013lka}. For \(M_U\gg2m_e\), Eq.~(\ref{BrUll}) for \(l=e\) reduces to:
\begin{equation}
    \Gamma(U \to e^+ e^-) \approx \frac{1}{3}\,\alpha\,\varepsilon^2\,M_U\,, \label{BrUll2}
\end{equation}
where in this case the corresponding proper lifetime of the dark photon is 
\begin{align}
    \tau_U &\;=\;\frac{\hbar}{\Gamma(U\to e^+e^-)}
    \approx \frac{3\,\hbar}{\alpha\,\varepsilon^2\,M_U},
    \nonumber\\
    &\approx 2.65\times10^{-16}\,\mathrm{s}
      \;\times\;
      \left(\frac{10^{-6}}{\varepsilon^2}\right)
      \left(\frac{1\ \mathrm{GeV}/c^2}{M_U}\right)\,.
\end{align}

For typical parameters (\(\varepsilon^2 \approx 10^{-6}\), \(M_U \approx 1\) GeV$/c^2$), the dark photon’s proper lifetime is  
$\tau_U \approx 10^{-16}\,\mathrm{s},$
so its decay length is \(c\tau_U \approx 10^{-8}\) m, i.e.\ only a few tens of nanometres.

The branching ratio $Br^{U\to e^+e^-}$ is define  as
\begin{eqnarray}
   Br^{U\to ee} &&= \frac{\Gamma_{U \rightarrow e^+e^-}}{\Gamma_{tot}^U} \label{Bree2},
\end{eqnarray}
where the total decay width of a dark photon is the sum of the partial decay widths 
to hadrons, $e^+e^-$, and $\mu^+\mu^-$ pairs,
\begin{eqnarray}
\Gamma_{tot}^U= \Gamma_{U\to hadr} + \Gamma_{U\to e^+e^-} + \Gamma_{U\to\mu^+\mu^-},
\end{eqnarray}
where $\Gamma_{U\to e^+e^-} $ and $ \Gamma_{U\to\mu^+\mu^-}$ are given in Eq. (\ref{BrUll}).
The hadronic decay width of a dark photon
is chosen such that $\Gamma_{U\to hadr} = R(\sqrt{s}=M_U)\Gamma_{U\to\mu^+\mu^-}$, where 
the factor $R(\sqrt{s}) = \sigma_{e^+e^-\rightarrow hadrons}$/$\sigma_{e^+e^-\rightarrow \mu^+\mu^-}$ \cite{Batell:2009di} 
\begin{equation}
\Gamma_{U \rightarrow \text { hadr }}=\frac{1}{3} \alpha \varepsilon^2 M_U \sqrt{1-\frac{4 m_\mu^2}{M_U^2}}\left(1+\frac{2 m_\mu^2}{M_U^2}\right) R\left(M_U\right).\label{Vtohadrons}
\end{equation}
\begin{figure}[t]
\includegraphics[width=0.99\linewidth]{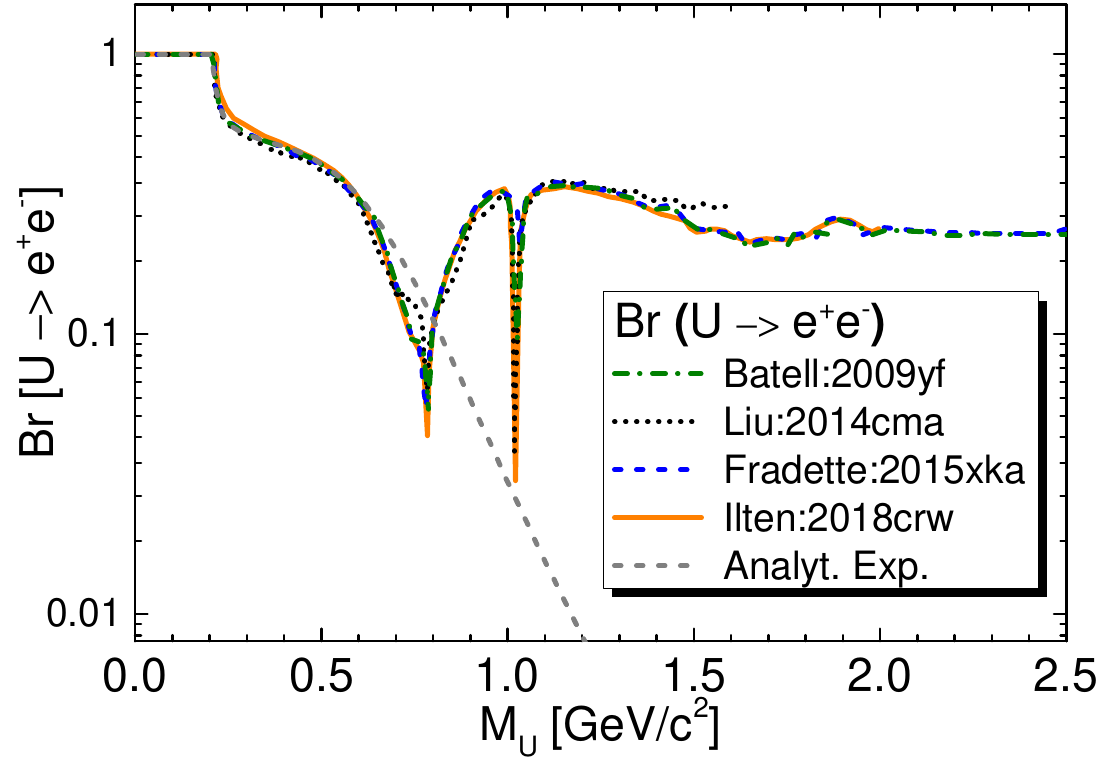}
  \caption{Branching ratio $Br(U\!\to\!e^+e^-)$ as a function of the dark‐photon mass $M_U$ (GeV/$c^2$).  Colored lines correspond to different theoretical predictions: the dispersive analysis of Batell \cite{Batell:2009yf} (green dot–dash), the tuned $R$–ratio model of Liu \cite{Liu:2014cma} (black dotted), the thermal‐production estimate of Fradette \cite{Fradette:2015xka} (blue dashed), and the vector‐meson dominance fit of Ilten \cite{Ilten:2018crw} (orange solid).  The gray dashed line shows the analytic low‐mass approximation from Eq.~(\ref{Bree}), which is only valid for $M_U\lesssim0.6\,$GeV/$c^2$.}
\label{BRfig}
\end{figure}
Now we evaluate the expression (\ref{Bree2})  using that $\Gamma_{U\to\mu^+\mu^-} = \Gamma_{U\to e^+e^-}$ 
(due to lepton universality for $M_U\gg 2m_\mu$) leading to
\begin{equation}\label{Bree}
Br^{U\to ee} =
\begin{cases} 
1, & M_U < 2m_\mu, \\[10pt]
\frac{1}{1 + \sqrt{1 - \frac{4m_\mu^2}{M_U^2}} \left( 1 + \frac{2m_\mu^2}{M_U}\right) \left(1 + R(M_U)\right)}, & M_U \geq 2m_\mu,
\end{cases}
\end{equation}
where $m_\mu$ is the muon mass and we have asummed $M_U \gg 2m_\mu$.

The analytic expression in Eq.~(\ref{Bree}), adopted from Ref.~\cite{Batell:2009yf} and also used in Ref.~\cite{Agakishiev:2013fwl}, provides a simple low–mass approximation for $Br(U\!\to\!e^+e^-)$ valid up to $M_U\lesssim0.6\,$GeV/$c^2$, where hadronic channels (e.g.\ $\pi^+\pi^-$, $K^+K^-$) remain closed and the total width is dominated by its leptonic modes. For $M_U>0.6\,$GeV/$c^2$, we extend the branching‐ratio prediction up to $M_U<2.5\,$GeV/$c^2$ using the vector‐meson–dominance fit of Ilten \cite{Ilten:2018crw}.

Fig. \ref{BRfig} displays $Br(U\to e^+e^-)$ versus $M_U$ (up to 2.5 GeV/$c^2$), comparing this analytic line (gray dashed) with four state‐of‐the‐art theoretical treatments: the dispersive analysis of Batell et al. \cite{Batell:2009yf}, the tuned $R$-ratio model of Liu \cite{Liu:2014cma}, the thermal‐production estimate of Fradette \cite{Fradette:2015xka}, and the updated vector meson dominance (VMD) parameterization of Ilten et al. \cite{Ilten:2018crw}. On the logarithmic vertical axis, all models converge to $Br \approx1$ for $M_U\lesssim 2m_\mu$, where hadronic decays are kinematically forbidden. As $M_U$ approaches the $\rho$, $\omega$, and $\phi$ resonance regions, the Batell and Liu treatments resolve the detailed dips and peaks at hadronic thresholds, Ilten’s VMD fit sharpens the narrow $\omega$ and $\phi$ features, and Fradette’s thermal corrections remain in close agreement outside the resonances. The gray dashed analytic line (Eq. (\ref{Bree})) falls off sharply above 0.6 GeV/$c^2$, illustrating the impact of neglected hadronic modes.

Having established the various predictions for the dark photon decay branching ratio and its dependence on $M_U$, we now proceed to implement these inputs within our transport approach.  
In the following section, we describe how dark photons are produced and subsequently decay into dileptons in the PHSD model.

\section{Dark photon production and dilepton decay channels  in the PHSD}\label{sec:DPproduction}

\subsection{Dark Photon Production}

In this study, we adopt the methodology developed in our prior papers ~\cite{Schmidt:2021hhs,Bratkovskaya:2022cch}. Alongside the previously considered Dalitz decays of $\pi^0, \eta $ and $\Delta$-resonances. Additionally, we consider further production channels for dark photons: direct decays of vector mesons, specifically $V = \rho, \phi, \omega$ as well as the  Dalitz decay of $\omega, \eta^{\prime}$, $K^+$ as well as 
$q\bar q$ annihilation.
\begin{eqnarray}
\pi^{\mathbf{0}}, \eta \rightarrow \gamma U, \label{decay1} \\
\Delta \rightarrow N U, \\
\rho, \boldsymbol{\phi}, \omega \rightarrow U, \\
\boldsymbol{\omega} \rightarrow \boldsymbol{\pi}^{\mathbf{0}} U,  \\
\eta^{\prime} \rightarrow \gamma U, \\
K^{+} \rightarrow \boldsymbol{\pi}^{+} U,\\
q\bar{q} \to U. \label{decay10}
\end{eqnarray}

Depending on the dark photon mass, the dominant production mechanisms are:
\begin{align}
    &\text{For: } M_U \lesssim m_\pi, & & \pi^{\mathbf{0}} \rightarrow \gamma U, \\
    &\text{For: } m_\pi \lesssim M_U \lesssim m_\eta, & & \eta \to \gamma U, \; \Delta \to N U, \\
    &\text{For: } m_\eta \lesssim M_U \lesssim m_\omega, & &\omega, \rho \to U,\\
    &\text{For: } m_\omega \lesssim M_U \lesssim m_\phi, & &\omega, \rho, \phi \to U,\\
    &\text{For: } M_U  \lesssim  m_\phi, & & \omega, \rho, \phi, \ \ q\bar{q} \to U.
\end{align}

Furthermore, rare modes such as  
\(K^+\to e^+\nu U\) and \(\Sigma^+\to p\,U\) are neglected, because their rates are already strongly limited by muon–anomalous–moment constraints and contribute negligibly in the mass window of interest~\cite{Pospelov:2008zw}. 
 A further channel $D^{*0}\to D^0U$ with $M_U<0.14$\,GeV$/c^2$ is likewise omitted~\cite{Ilten:2015hya}. Moreover, the contributions from correlated charm decays, primary Drell-Yan processes ($q\bar{q} \rightarrow U$) and electromagnetic bremsstrahlung ($e^-Z \rightarrow e^-Z U$)
as proposed in Refs.~\cite{Fabbrichesi:2020wbt,Alexander:2016aln,Berlin:2018pwi} respectively,  are not  included in our study.

\subsection{Dark Photon Yield}
The  number of dark photon decays of mass $M_U$ can be evaluated as the sum of all possible contributions 
for a given mass $M_U$:
\begin{eqnarray}
N_{U} &= &  N_{\pi\to\gamma U}+ ... +  N_{q\bar{q}\to  U},\\&= &
\sum_{h} N_{h \to  U X} ,\label{NUee}
\end{eqnarray}
\noindent
where $h$ and $X$ involve all channels from Eq. (\ref{decay1})-(\ref{decay10}): $X = \gamma$ for $h = \pi, \eta, \eta^{\prime}, \omega$, $X = N$ for $h = \Delta$,  $X = \pi$ for $h = \omega, K^+$ and  $h=\omega,\rho, \phi $ and $q\bar{q}$ for direct decays without $X$.
We assume that the width of the dark photon is zero (or very small), i.e., it contributes only 
to a single $dM$ bin of the dilepton spectra from SM sources.

\subsubsection{Dalitz Decays of $\pi^0, \eta,\eta^{\prime} $ and $ \omega$ }

The yield of dark photons of mass $M_U$ themselves can be
estimated from the coupling to $\pi^0$, $\eta, \eta^{\prime}$ and $ \omega$ decays to the virtual photons \cite{Agakishiev:2013fwl}:
\begin{eqnarray}
&& N_{m\to \gamma U} = N_m Br_{m\to \gamma \gamma}  \cdot
   \frac{\Gamma_{m\rightarrow\gamma U}}{\Gamma_{m\rightarrow\gamma\gamma}}, \ \ m=\pi^0, \eta, \eta^{\prime},  \label{mNU1} \\
&& N_{\omega\to \pi^0 U} = N_\omega Br_{\omega\to \pi^0 \gamma}  \cdot    
   \frac{\Gamma_{\omega\to \pi^0 U}}{\Gamma_{\omega\to \pi^0 \gamma}}. \label{mNU3}
\end{eqnarray}

These ratios of the partial widths for the Dalitz decays of $\pi^0$, $\eta$, 
$\eta^{\prime}$ and $\omega$ mesons  to dark photons
can be evaluated following Refs. \cite{Batell:2009yf,Batell:2009di,Gorbunov:2024nph,Berlin:2018pwi}:
\begin{eqnarray}
  &&  \frac{\Gamma_{m\rightarrow\gamma U}}{\Gamma_{m\rightarrow\gamma\gamma}} = 2\varepsilon^2|F_m(q^2 = M_U^2)|\frac{\lambda^{3/2}(m_m^2, m_\gamma^2, M_U^2)}{\lambda^{3/2}(m_m^2, m_\gamma^2, m_\gamma^2)},\label{GgU}    \\
 && \frac{\Gamma_{\omega \to \pi^0 U}}{\Gamma_{\omega \to \pi^0 \gamma}} = \varepsilon^2 \frac{\lambda^{3/2}(m_\omega^2, m_{\pi^0}^2, M_U^2)}{\lambda^{3/2}(m_\omega^2, m_{\pi^0}^2, m_\gamma^2)}.
\end{eqnarray}

 Note that the factor 2 in Eq. (\ref{GgU}) arises because in the $m \to \gamma \gamma$ decay there are two identical photons, yielding two equivalent diagrams when one photon is replaced by the dark photon. Here  $\varepsilon^2$ is the kinetic mixing parameter and $\lambda$ is the triangle function 
$\lambda(x,y,z)=(x-(y-z)^2)(x-(y+z)^2)$ from the expression of particle 3-momentum. Since $m_\gamma=0$,
one obtains:
\begin{eqnarray}
 &&   \frac{\lambda^{3/2}(m_m^2, 0, M_U^2)}{\lambda^{3/2}(m_m^2, 0, 0)} = \left(  1 - \frac{M_U^2}{m_m^2}\right)^3, \label{lam}     \\
&& \frac{\lambda^{3/2}(m_\omega^2, m_{\pi^0}^2, M_U^2)}{\lambda^{3/2}(m_\omega^2, m_{\pi^0}^2, 0)} = \frac{1}{\left(m_\omega^2-m_\pi^2\right)^3} \times \nonumber \\
&&  \left[\left(m_\omega^2-\left(M_U+m_\pi\right)^2\right)\left(m_\omega^2-\left(M_U-m_\pi\right)^2\right)\right]^{3 / 2},
\end{eqnarray}
here, $M_U$ is the dark photon mass, $F_m$ are the electromagnetic transition formfactors for $\pi^0$ and $\eta$; they are taken as  
in our previous studies \cite{Bratkovskaya:2007jk,Bratkovskaya:2013vx}.

\subsubsection{Dalitz Decays of $\Delta$}

The dark photon production by the $\Delta$ Dalitz decay $\Delta \to NU$ has been proposed in Ref. \cite{Batell:2009di},
\begin{equation}
 N_{\Delta\to N U} = N_\Delta Br_{\Delta\to N \gamma}  \cdot    
   \frac{\Gamma_{\Delta\rightarrow NU}}{\Gamma_{\Delta\rightarrow N\gamma}}. \label{mNU2}
\end{equation}
For the evaluation of the partial decay widths of a broad $\Delta$  resonance, one has to take into account
the $\Delta$ spectral function $A(M_\Delta)$ as used also in the HADES study \cite{Agakishiev:2013fwl}:
\begin{eqnarray}
&&   \frac{\Gamma_{\Delta\rightarrow NU}}{\Gamma_{\Delta\rightarrow N\gamma}}  \label{DNU} \\
  &&= \varepsilon^2\int A(M_\Delta)|F_\Delta(M_U^2)|\frac{\lambda^{3/2}(M_\Delta^2, m_N^2, M_U^2)}{\lambda^{3/2}(M_\Delta^2, m_N^2, m_\gamma^2)} dM_\Delta, \nonumber   
\end{eqnarray}
where  $M_\Delta$ is the mass of the $\Delta$ resonance distributed according to the spectral function $A(M_\Delta)$,
$m_N$ the mass of the remaining nucleon. 
Following Ref. \cite{Agakishiev:2013fwl} we adopted $|F_\Delta(M_U^2)| = 1$ for this study
since an experimental formfactor is unknown.

In the PHSD the spectral function of a $\Delta$ resonance of mass $M_\Delta$
is taken in the relativistic Breit-Wigner form \cite{Bratkovskaya:2013vx}:
\begin{eqnarray}
\footnotesize
A_\Delta(M_\Delta) = C_1\cdot {2\over \pi} \ {M_\Delta^2 \Gamma_\Delta^{tot}(M_\Delta)
\over (M_\Delta^2-M_{\Delta 0}^2)^2 + (M_\Delta {\Gamma_\Delta^{tot}(M_\Delta)})^2}.
\label{spfunD}
\end{eqnarray}
with $M_{\Delta 0}$ being the pole mass of the $\Delta$.
The factor $C_1$ is fixed by the normalization condition:
\begin{eqnarray}
\int_{M_{min}}^{M_{lim}} A_\Delta(M_\Delta) dM_\Delta =1,
\label{SFnorma}\end{eqnarray} where $M_{lim}=2$~GeV$/c^2$ is chosen as
an upper limit for the numerical integration. The lower limit for
the vacuum spectral function corresponds to the nucleon-pion decay,
$M_{min}=m_\pi+m_N$. In  $NN$ collisions the $\Delta$'s can be populated up to the $M_{max}=\sqrt{s}-m_N$ and hence
the available part of the spectral function depends on the collision energy.
\begin{figure*}[t]
\includegraphics[width=0.49\linewidth]{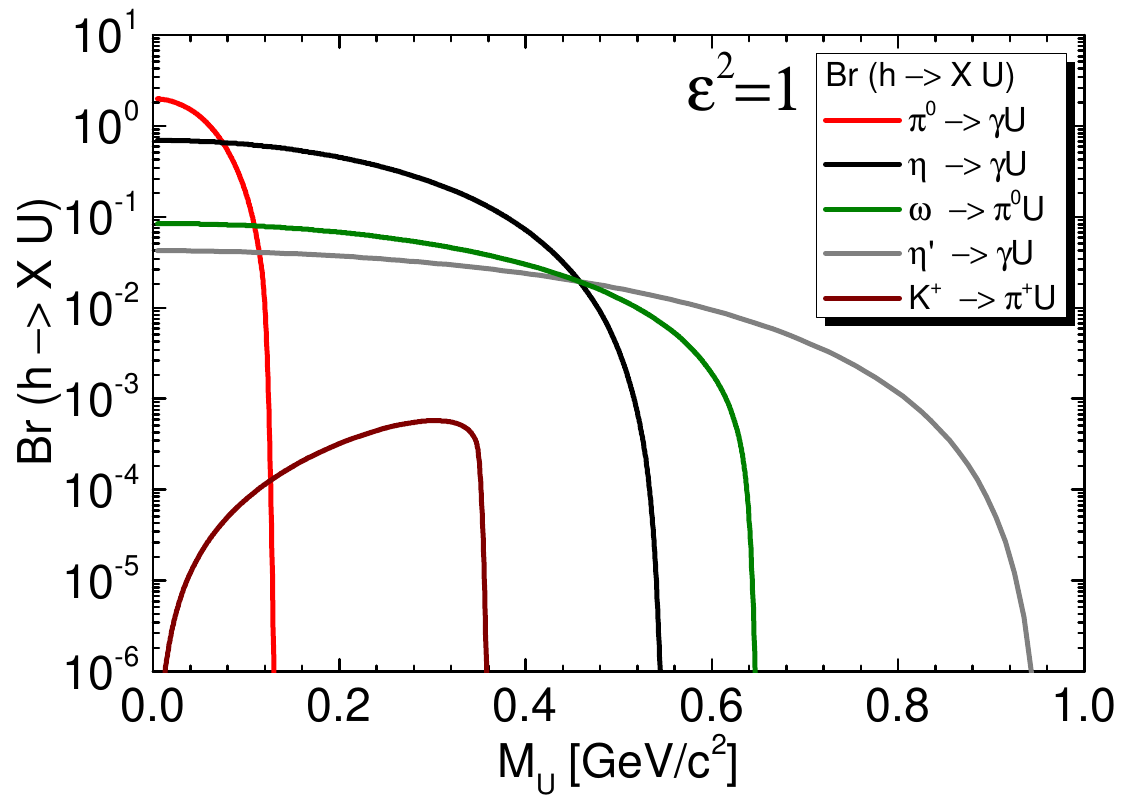}
\includegraphics[width=0.49\linewidth]{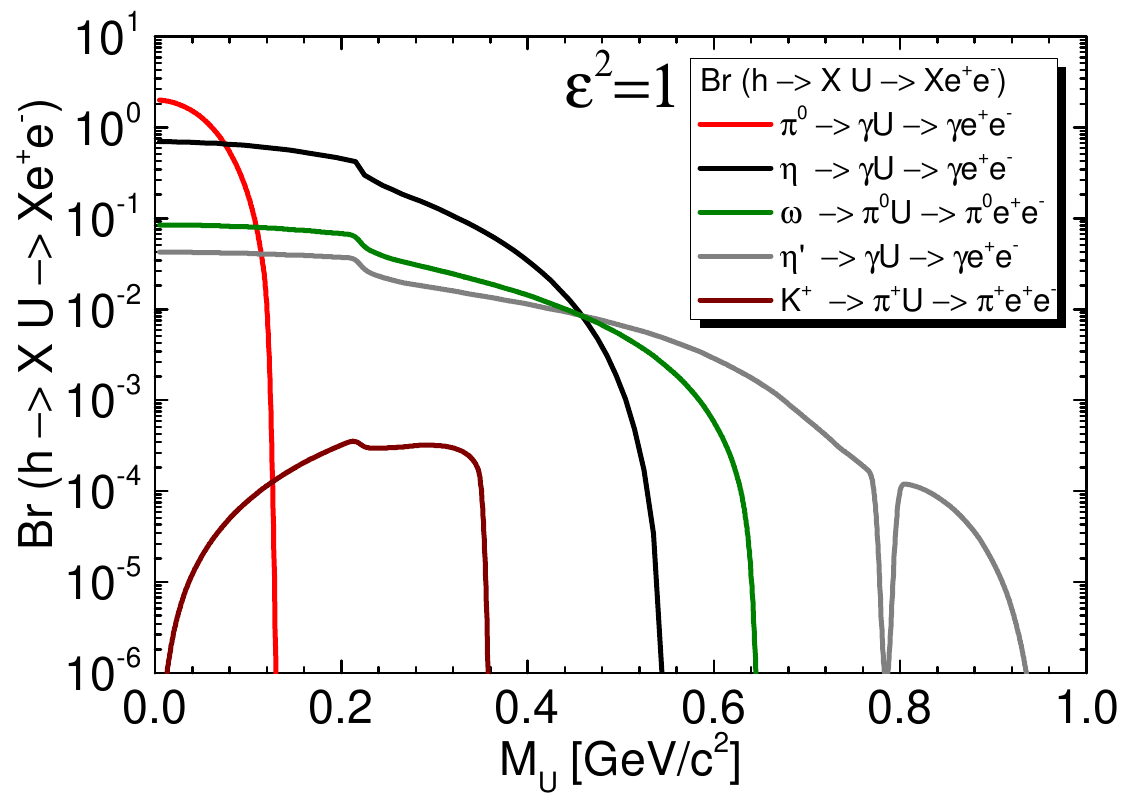}
\caption{Branching ratios as a function of the  dark photon mass for meson decays to dark photons ($U$-bosons) and subsequent dilepton decays. The left panel shows the branching ratios for mesons decaying into dark photons: $\pi^0 \to \gamma U$, $\eta \to \gamma U$, $\omega \to \pi^0 U$, $\eta' \to \gamma U$, and $K^+ \to \pi^+ U$. The right panel shows the branching ratios for meson decays into dark photons and subsequent dilepton decays: $\pi^0 \to \gamma U \to \gamma e^+ e^-$, $\eta \to \gamma U \to \gamma e^+ e^-$, $\omega \to \pi^0 U \to \pi^0 e^+ e^-$, $\eta' \to \gamma U \to \gamma e^+ e^-$, and $K^+ \to \pi^+ U \to \pi^+ e^+ e^-$. In both panels, the lines represent different meson decay channels. The kinetic mixing parameter $\varepsilon^2$ is set to 1. }
\label{BRfig2}
\end{figure*}
We recall that for the total decay width of the $\Delta$ resonance $\Gamma_\Delta^{tot} (M_\Delta)$ in the PHSD  
we adopt the "Monitz model" \cite{Koch:1983tf} (cf. also Ref. \cite{Wolf:1990ur}):
\begin{eqnarray}
&&\Gamma_\Delta^{tot} (M_\Delta)= \Gamma_R {M_{\Delta 0} \over M_\Delta}
       \cdot \left(q\over q_r\right)^3 \cdot F^2(q), \label{WidthDel}\\
&& q^2={(M_\Delta^2 -(m_N+m_\pi)^2)(M_\Delta^2 -(m_N-m_\pi)^2)
       \over 4 M_\Delta^2}, \nonumber \\
&&  \Gamma_R = 0.11 {\ \rm GeV}, \ \ M_{\Delta 0} = 1.232 {\ \rm GeV/c^2}; \nonumber \\
&& F(q) ={\beta_r^2 +q_r^2 \over \beta_r^2 +q^2}, \label{WidthDel1} \\
&&  q_r^2 = 0.051936, \ \ \beta_r^2 = 0.09.\nonumber
\end{eqnarray}
We note that when accounting for the mass-dependent total width of the $\Delta$ resonance  our calculation for the 
dark photon production by the $\Delta$ Dalitz decay differs from the evaluation in Ref. \cite{Agakishiev:2013fwl}
where a constant total width of the $\Delta$ has been used. As discussed in 
Ref. \cite{Bratkovskaya:2013vx} (see Section VI), the shape of the spectral function strongly depends on the
actual form of $\Gamma_\Delta^{tot} (M_\Delta)$.
\vspace{-0.5cm}
\subsubsection{Direct Decays of $\omega,\rho, \phi$.}

For the dark photon production of  $h=\rho, \phi, \omega$ in Ref. ~\cite{Batell:2009di},
\begin{eqnarray}
 N_{V \to  U} = N_V \cdot Br_{V \to U}
, \quad \quad V=\rho, \phi, \omega,\label{mNU6}
\end{eqnarray}
where
every virtual photon $\gamma^*$ is replaced by a dark photon with electromagnetic strenght  $\alpha^{\prime}=\alpha \times \varepsilon^2$ following the dark photon model and taking $Br_{V \to U}=1$. 

\subsubsection{Dalitz Decays of $K^+$}
In the case of $K^+$ decay we follow the same procedure, but the branching ratio is known and taken from Ref. ~\cite{Pospelov:2008zw}, then, the number of dark photons produced in this decay is
\begin{eqnarray}
&& N_{K^+\to \pi^+ U} = N_K^+ \cdot Br_{k^+ \to \pi^+ U}, \label{mNU5}
\end{eqnarray}
where
\begin{align}
& B r\left(K^{+} \rightarrow \pi^{+} U\right)=\frac{\alpha \varepsilon^2}{2^{10} \pi^2 \Gamma_T(K)} \frac{M_U}{m_K} W\left(M_U,m_K\right) \times \nonumber \\
& 
\hspace{5cm} \lambda^{1 / 2}\left(M_U^2, m_K^2, m_\pi^2\right),
\end{align}
where
$W=10^{-12}(3+6M_U^2/m_K^2)$ and $\Gamma_{T}(K)$ is the total width of $K^+$.
\vspace{-0.5cm}
\subsubsection{$q\bar q$ annihilation to a dark photon ($q\bar{q} \to U$) }

In the case of $q\bar q$ annihilation to dark photon  ($q\bar{q} \to U$) we use the following relation to rescale it to dark photons multiplying by the kinetic mixing parameter, where the number of $e^+e^-$ pairs is computed following the same procedure of Ref. \cite{SHiP:2020vbd,Berlin:2018pwi} 
\begin{eqnarray} 
&& N^{U \to e^+e^-}_{q\bar{q}} = \varepsilon^2 \cdot N_{q\bar{q} \to  e^+e^-}, \label{NUqqee} 
\end{eqnarray}
where the dilepton yield $N_{q\bar{q} \to  e^+e^-}$ is computed in PHSD.

We note that in spite that the dark photon is considered as an on-shell particle with mass $M_U$ such process $q\bar{q} \to U$ is kinematically allowed due to the off-shell nature of the DQPM partons. Such a process would not be allowed in the case of massless pQCD (anti-)quarks  without a production of a second particle to obey energy-momentum conservation.

\subsection{Total Dark Photon Dilepton Yield}

The total dilepton yield from a dark photon decay of mass $M_U$ can be evaluated as for a given mass $M_U$ as the total sum of all channels (\ref{decay1}-\ref{decay10}):
\begin{eqnarray}
 N^{U\to e^+e^-}  
 &=&  Br^{U\to e^+e^-} 
 \sum_{h}
 N_{h \to  U X}+ \nonumber\\
&&  N_{q\bar{q}}^{U\to e^+e^-},\label{yieldtotal}
\end{eqnarray}
\noindent
where $Br^{U\to e^+e^-}$  multiplies to the number of total dark photons produced in each channel, except for the decay $q\bar{q} \to U$ because it is calculated by rescaling to the kinemtic mixing parameter (Eq. (\ref{NUqqee})).

The left panel of Fig. \ref{BRfig2} shows the  branching fractions $Br(h\to X\,U)$ of light mesons for $h=\pi^0,\eta,\omega,\eta',K^+$ into the dark photon $U$  versus the dark photon mass $M_U$ assuming the kinetic mixing $\varepsilon^2=1$. The $\pi^0\to\gamma U$ channel (red) steeply drops at $M_U\approx m_{\pi^0}$, while the heavier mesons exhibit thresholds at their respective mass gaps ($\eta$ at $\approx 0.55\,$GeV$/c^2$, $\omega$ at $\approx 0.65\,$GeV$/c^2$, $\eta'$ at $\approx 0.96\,$GeV$/c^2$, and $K^+$ at $\approx 0.36\,$GeV$/c^2$). 

The right panel  of Fig. \ref{BRfig2}  displays the effective branching ratios for the full decay chain $h\to X\,U\to X\,e^+e^-$. The shapes mirror those in the left plot, but are suppressed by the leptonic decay $Br(U\to e^+e^-)$, yielding in  small reduction. Notably, the $\eta\to\gamma U\to\gamma e^+e^-$ (black) line remains dominant in the intermediate mass region, while the $\pi^0\to\gamma U\to\gamma e^+e^-$ channel (red) dominates only at very low $M_U$. This comparison highlights which meson decays and mass ranges provide the most promising $e^+e^-$ signals for dark photon searches.

Having defined the total dilepton yield from all dark photon decay channels (Eq.(\ref{yieldtotal})) and illustrated the branching-ratio behavior in Fig. \ref{BRfig2}, we now proceed to compare these predictions with the simulated dilepton spectra in order to extract theoretical constraints on the kinetic mixing parameter $\varepsilon^2(M_U)$.

\section{Results for the dilepton spectra from dark photon decays and constraints on $\varepsilon^2 (M_U)$}\label{sec:results}

\subsection{How to get theoretical constraints on  $\varepsilon^2 (M_U)$ }
 Given that both the kinetic mixing parameter, $\varepsilon^2$, and the mass of the dark photon are not yet determined, we utilize the following approach to impose constraints on $\varepsilon^2(M_U)$. For each dilepton mass bin $dM$, chosen as 10 MeV in our simulations, we compute the integrated dilepton yield from dark photons with masses within the range $[M_U, M_U + dM]$ using Eq. (\ref{NUee}) and normalize it by the bin width $dM$. The resulting dilepton yield per mass bin $dM$ is denoted as $dN^{sumU}/dM$, representing the sum of all kinematically accessible dilepton contributions from dark photons produced via the dark photon production channel $h \to U X$.

Assuming that $\varepsilon^2$ remains constant across $dM$, we express the dilepton yield as 
\begin{equation}\label{dNdmep1}
\dfrac{dN}{dM}^{sumU} = \varepsilon^2 \cdot \dfrac{dN_{\varepsilon=1}^{sumU}}{dM},
\end{equation}
where $dN^{sumU}_{\varepsilon=1}/dM$ 
represents the dilepton yield computed with $\varepsilon = 1$.
The total yield from all possible dilepton sources, including both SM channels and dark photon decays, can then be written as:
\begin{eqnarray}
  \frac{dN}{dM}^{total} &=& \frac{dN}{dM}^{sum SM} + \frac{dN}{dM}^{sum U} \\ &=&
    \frac{dN}{dM}^{sum SM} + \varepsilon^2 \cdot \frac{dN_{\varepsilon=1}^{sum U}}{dM}.
\label{dNdMepsil}
\end{eqnarray}
\begin{figure}[t]
\includegraphics[width=0.99\linewidth]{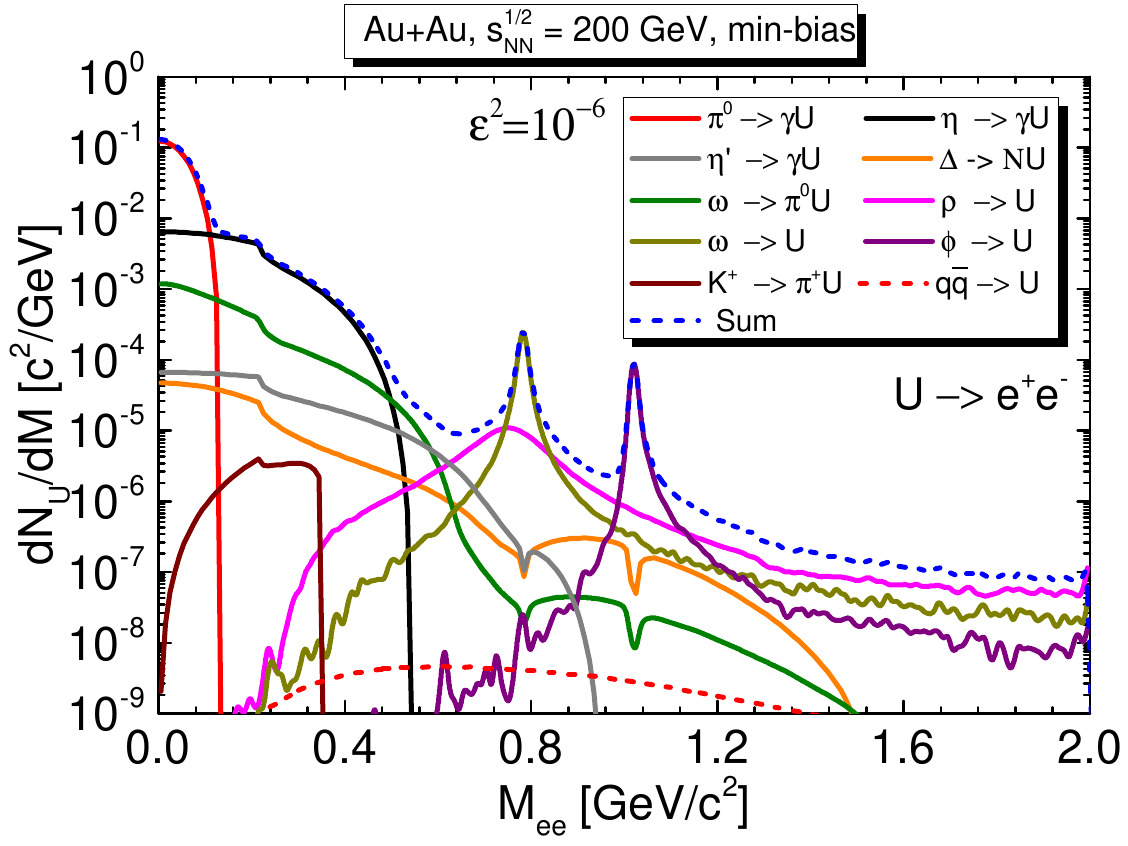}
\caption{Dark photon yield spectra
    $\mathrm{d}N_{U}/\mathrm{d}M$ 
    as a function of the invariant mass $M_{ee}$ (GeV/$c^2$) for minimum-bias
    Au+Au collisions at $\sqrt{s_{_{NN}}}=200\,$GeV, assuming a constant kinetic
    mixing parameter $\varepsilon^2=10^{-6}$. Colored lines indicate the
    contributions from individual production channels:
    $\pi^0\to\gamma U$ (light red),
    $\eta\to\gamma U$ (gray),
    $\omega\to\pi^0 U$ (green),
    $\eta'\to\gamma U$ (dark green),
    $\phi\to\eta U$ (magenta),
    $K^+\to\pi^+ U$ (brown),
    $\Delta\to N U$ (orange),
    $q\bar q\to U$ (dashed red), and the blue dashed line is the total sum.  }
\label{yieldAu200}
\end{figure}

\begin{figure*}[!]
\centerline{
\includegraphics[width=8.1 cm]{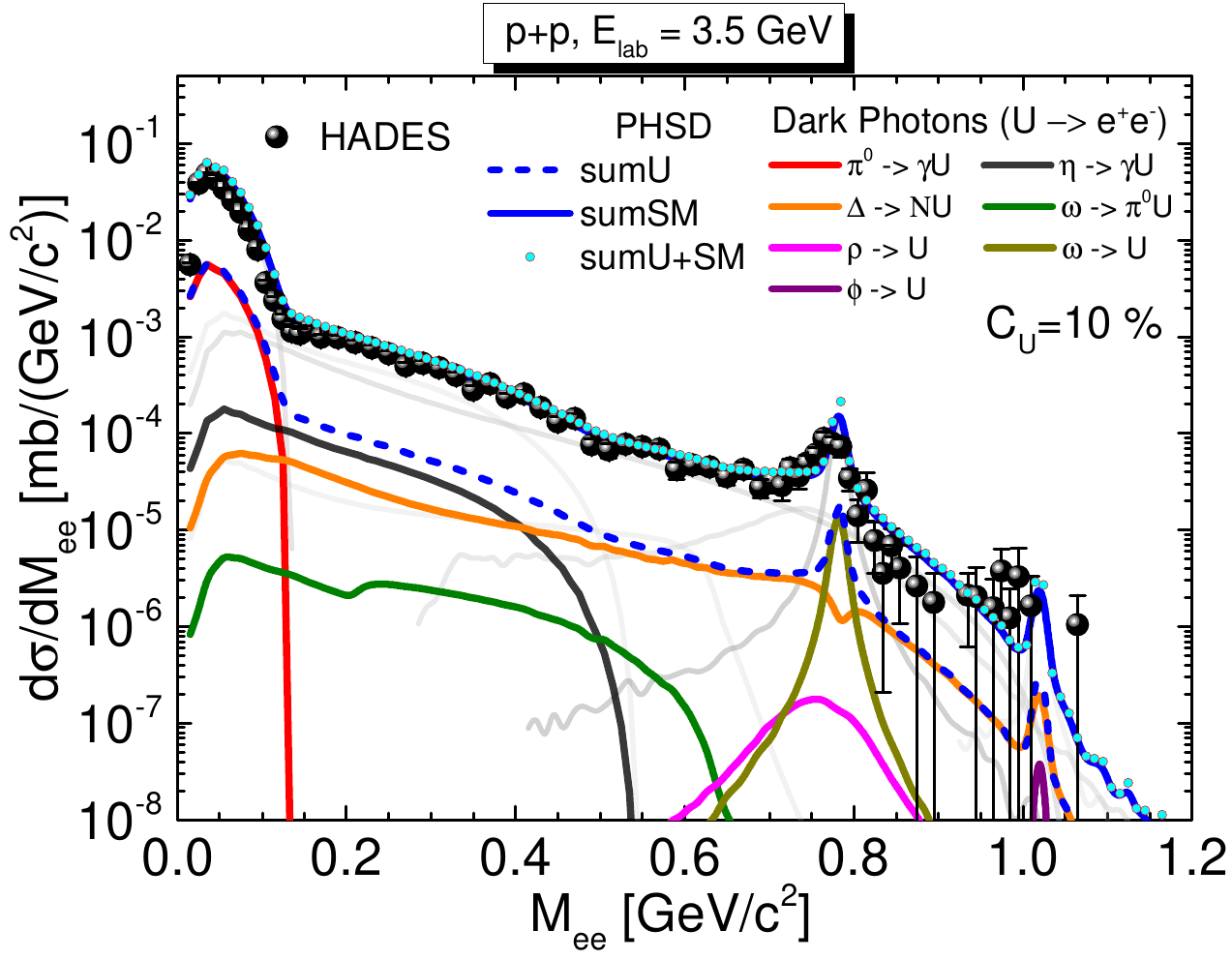}\hspace*{5mm}
\includegraphics[width=8.1 cm]{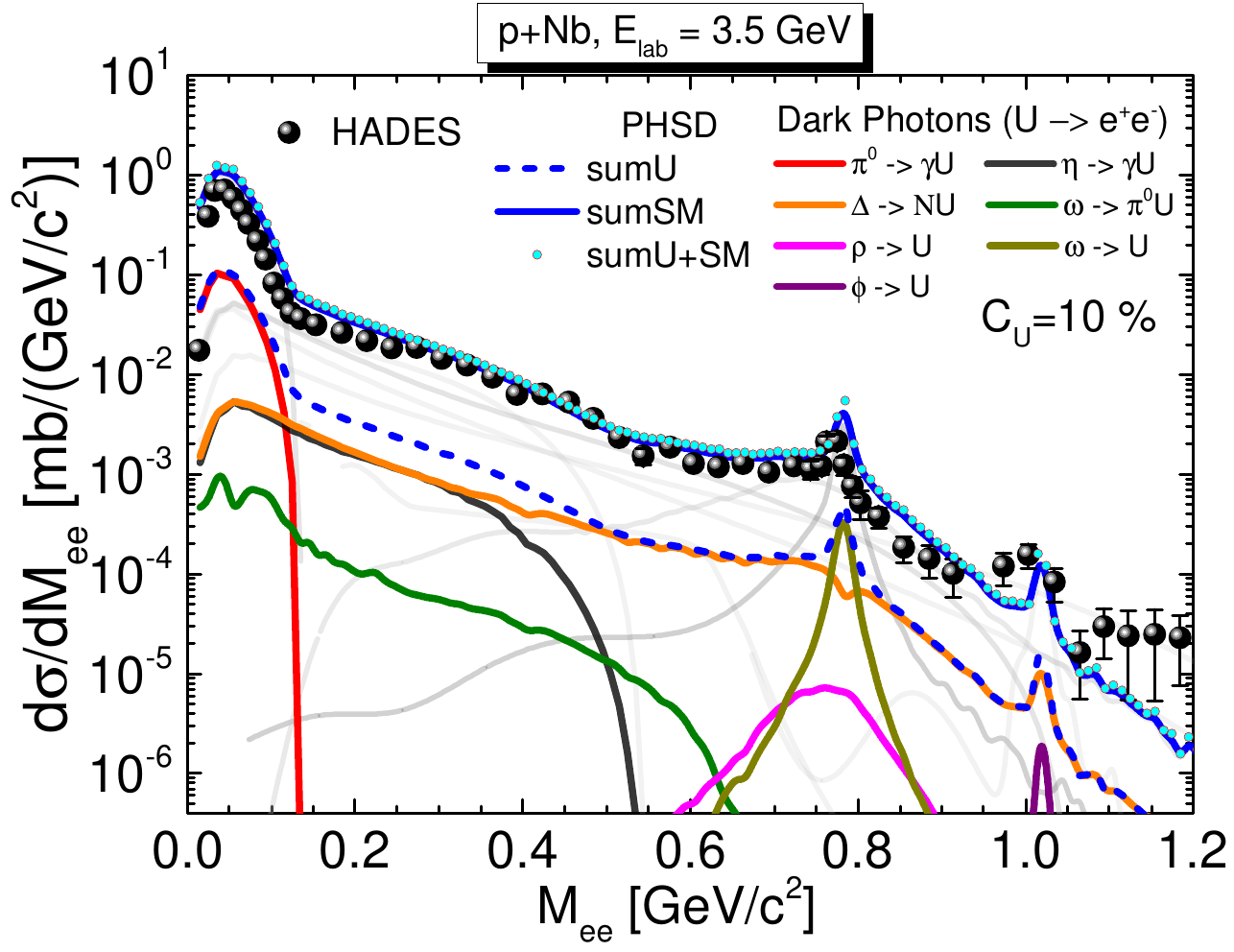}}
\centerline{
\includegraphics[width=8.1 cm]{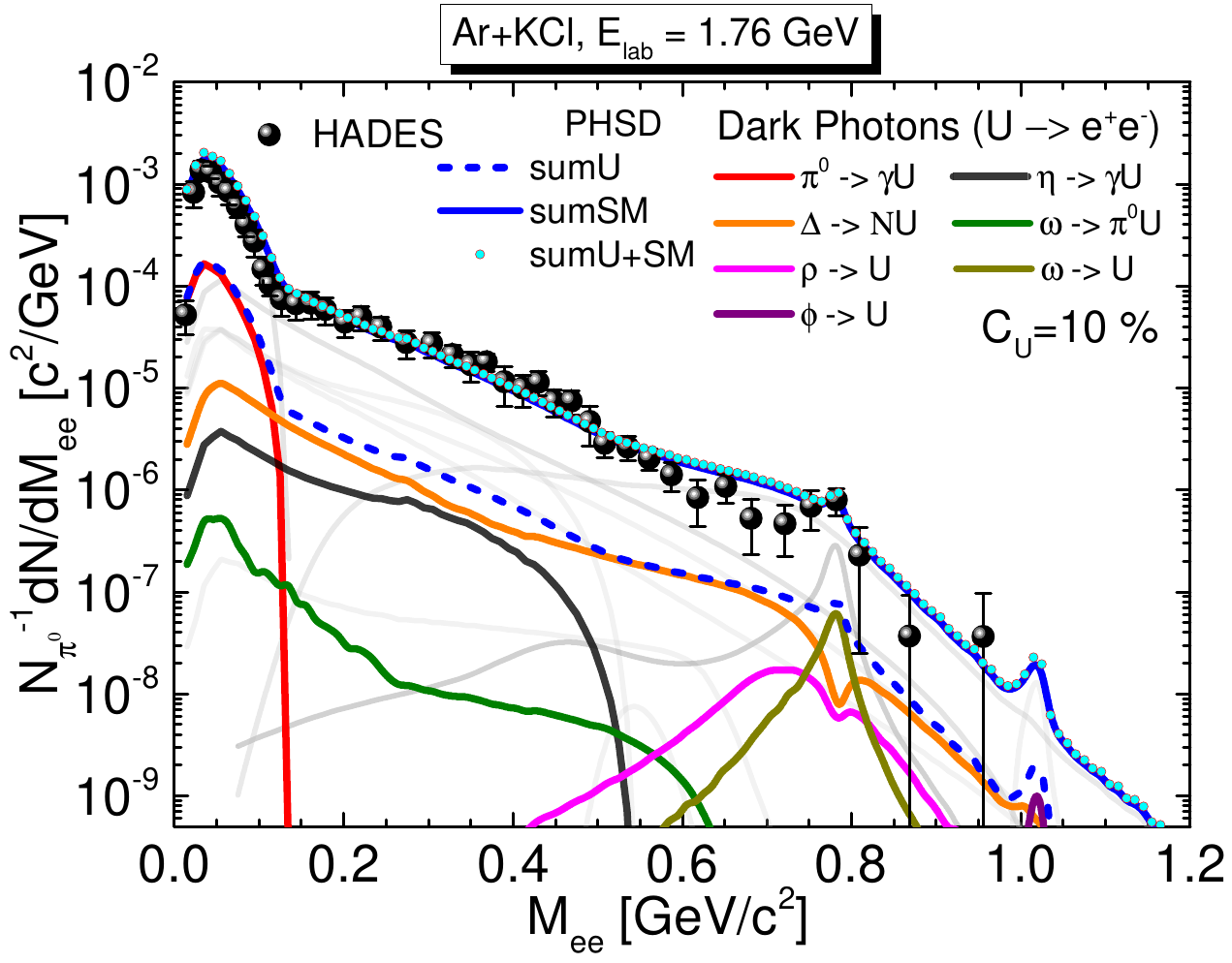}\hspace*{5mm}
\includegraphics[width=8.1 cm]{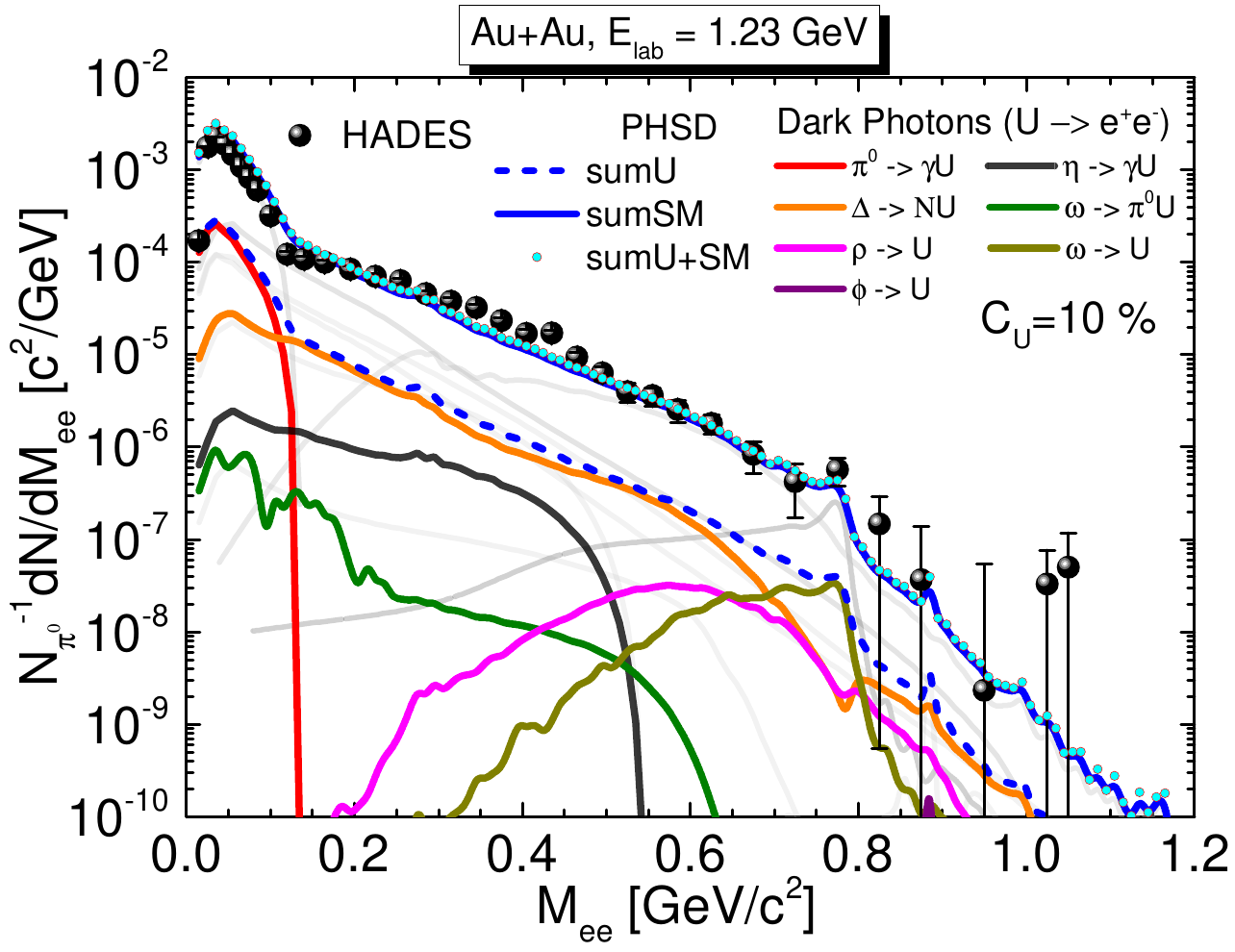}}
\caption{The PHSD simulations of the differential cross section $d\sigma/dM_{ee}$ for $e^+e^-$ production in $p+p$ (top left panel) and $p+Nb$ collisions (top right panel) at a beam energy of 3.5 AGeV. Additionally, the calculations provide the mass differential dilepton spectra $dN/dM$, normalized to the $\pi^0$ multiplicity, for $Ar+KCl$ collisions at 1.76 AGeV (bottom left panel) and for $Au+Au$ collisions at 1.23 AGeV (bottom right panel). These results are compared with experimental data from the HADES Collaboration.
The light gray lines denote the SM channels included in the PHSD calculations.
The HADES measurements are represented by solid dots: the $p+p$ data are taken from Ref. \cite{HADES:2011ab}, $p+Nb$ results from Refs. \cite{HADES:2011jqb,Agakishiev:2012vj}, $Ar+KCl$ data from Ref. \cite{Agakishiev:2011vf}, and $Au+Au$ data from Ref. \cite{HADES:2019auv}. The various SM contributions to dilepton production in PHSD are illustrated by individual colored lines, with the specific channels indicated in the legend.
Dileptons arising from $U \rightarrow e^+e^-$ processes, allowing for a 10\% surplus over the total SM yield, are included.
Dark photon contributions are categorized by their sources: Dalitz decays of $\pi^0$ (red), $\eta$ (black), $\Delta$ resonances (orange), $\omega$ (olive) and direct vector meson decays $\rho, \omega, \phi$ (magenta, dark yellow, purple) respectively. The combined yield from these decays is shown as a dashed blue line, while the total dilepton yield including both SM and dark photon contributions are represented by cyan dots.
We used the HADES acceptance criteria, incorporating its mass and momentum resolution. }
\label{M_Hades}
\end{figure*}
We can set constraints on $\varepsilon^2(M_U)$ by requiring that the total dilepton yield, $dN^{total}/dM$, does not exceed the Standard Model contribution by more than a specified fraction $C_U$ for each mass bin $dM$. The parameter $C_U$ defines the maximum permissible increase in a dilepton yield from dark photons compared to the Standard Model yield (for instance, setting $C_U = 0.1$ means that dark photons can contribute up to an additional 10\% to the yield observed from Standard Model processes). This condition can be formulated as
\begin{eqnarray}
   \frac{dN}{dM}^{sumU} = C_U \cdot  \frac{dN}{dM}^{sum SM}.
\end{eqnarray}
By combining the previous equation with Eq. (\ref{dNdmep1}), it is possible to express the kinetic mixing parameter $\varepsilon^2$ for a given $M_U$ as:
\begin{eqnarray}
    \varepsilon^2 (M_U) = C_U \cdot  \left. { \left(\frac{dN}{dM}^{sumSM} \right)} \right/
    {\left(\frac{dN_{\varepsilon=1}^{sum U}}{dM} \right)}.
\label{epsM}
\end{eqnarray}

\begin{figure*}[!]
    \centering
     \includegraphics[width=0.49\linewidth]{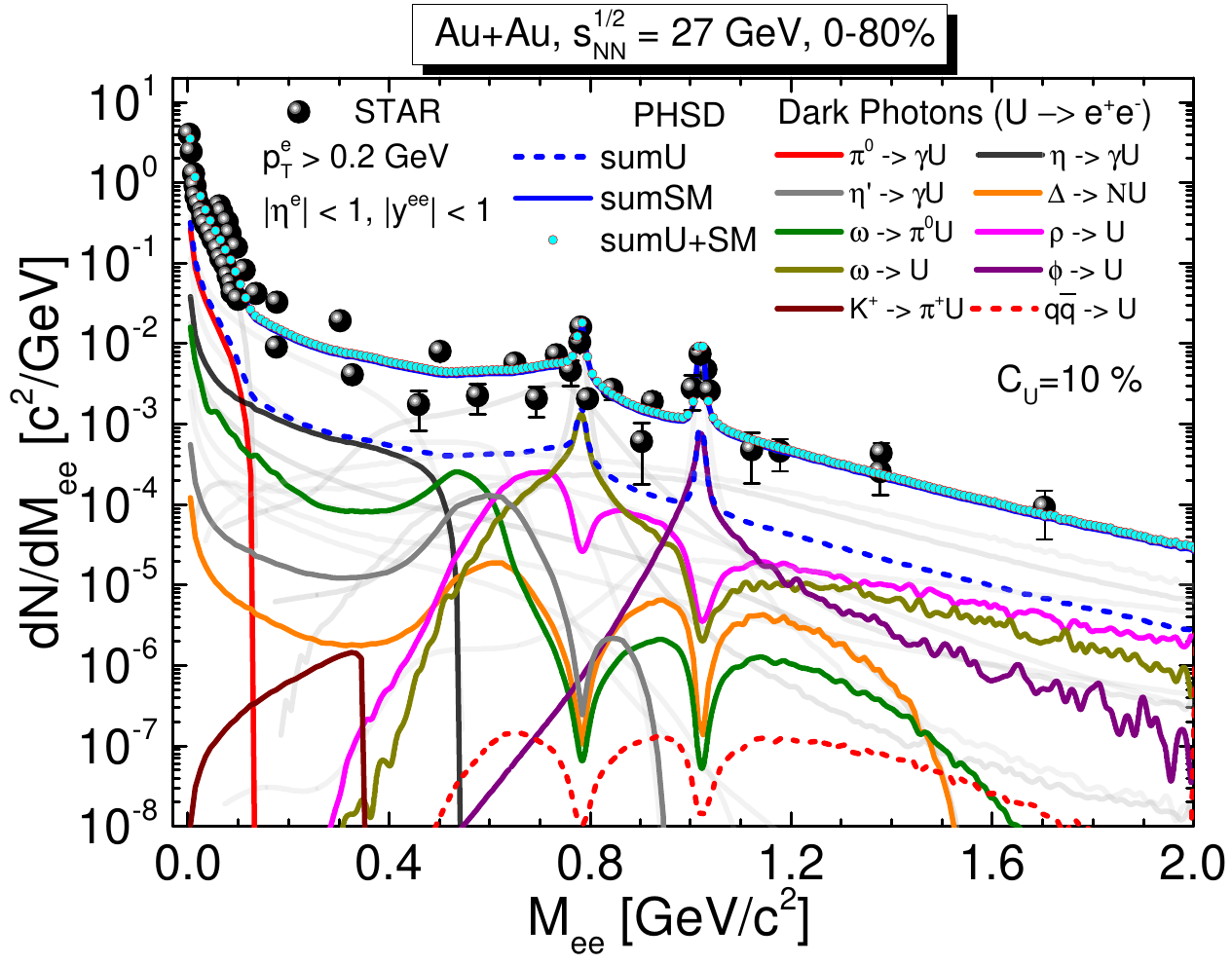}
     \includegraphics[width=0.49\linewidth]{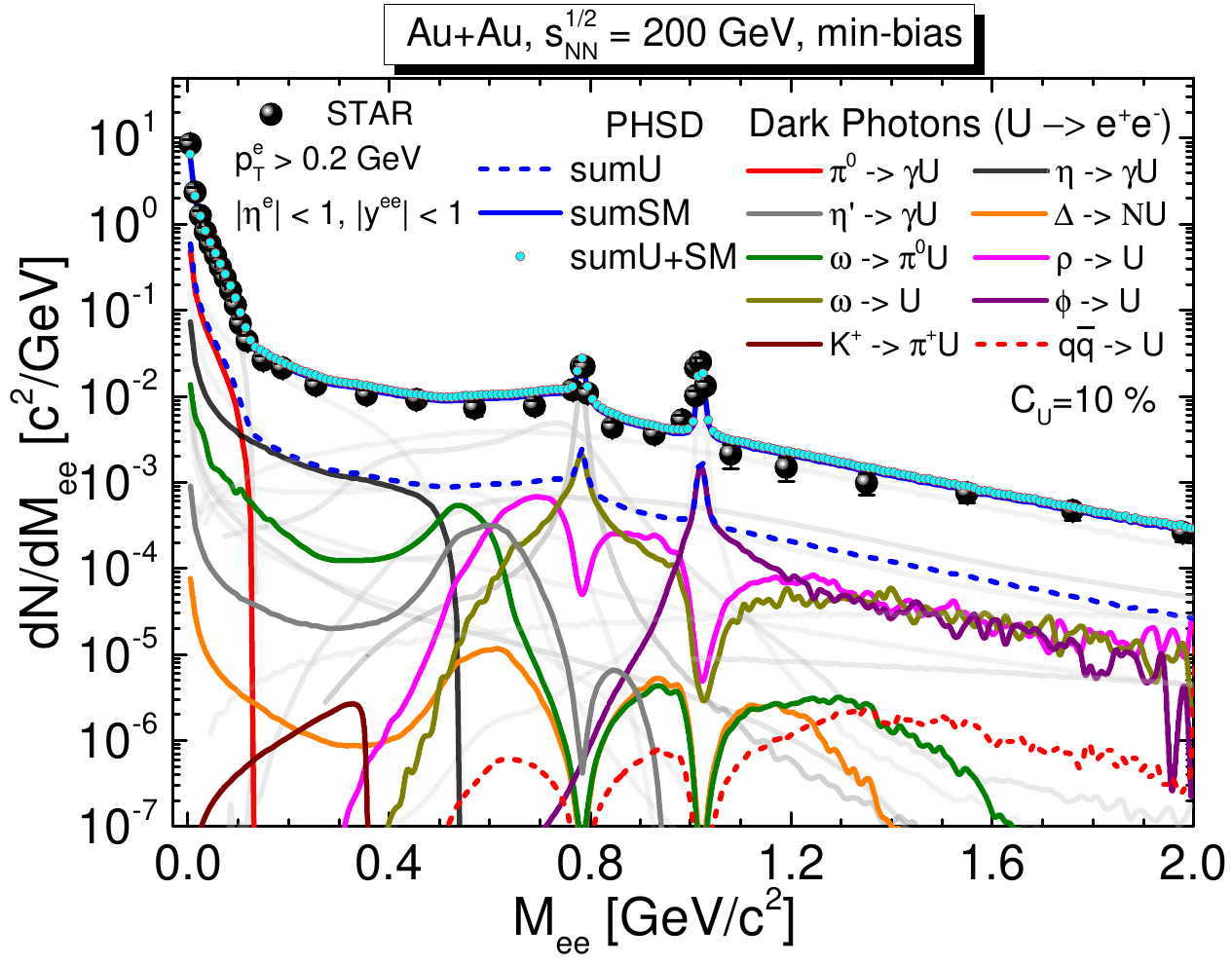}
    \caption{
The invariant mass spectra of dileptons produced in Au+Au collisions at $\sqrt{s_{NN}} = 27$ GeV (left panel) and $200$ GeV (right panel) are calculated using PHSD and compared to STAR experimental data  ~\cite{Han:2024nzr,STAR:2024bpc} and   ~\cite{STAR:2015tnn} respectively. The total dilepton yields predicted by the PHSD model are represented by solid blue lines, while the individual contributions from different production channels are detailed in the legends.
The light gray lines denote the SM channels included in the PHSD calculations. 
Dark photon contributions are categorized by their sources: Dalitz decays of $\pi^0$ (red), $\eta$ (black), $\eta^{\prime}$  (gray), $\Delta$ resonances (orange), $\omega$ (olive), direct vector meson decays  of $\rho, \omega, \phi$ (magenta, dark yellow, purple) respectively, $K^+ $ decay (brown) and $q\bar{q}$ annihilation (dashed red).
In addition, the solid lines reflect the inclusion of $U \rightarrow e^+e^-$ decays, allowing for a 10\% surplus over the total SM yield.
The combined yield from these decays is shown as a dashed blue line, while the total dilepton yield including both SM and dark photon contributions are represented by cyan dots.
The STAR experimental data points are depicted as solid black dots.
To facilitate a direct comparison, the theoretical calculations are adjusted to align with the STAR acceptance criteria, incorporating its mass and momentum resolution.    }
    \label{RHIC_AuAu}
\end{figure*}
\begin{figure*}[!]
    \centering
     \includegraphics[width=0.49\linewidth]{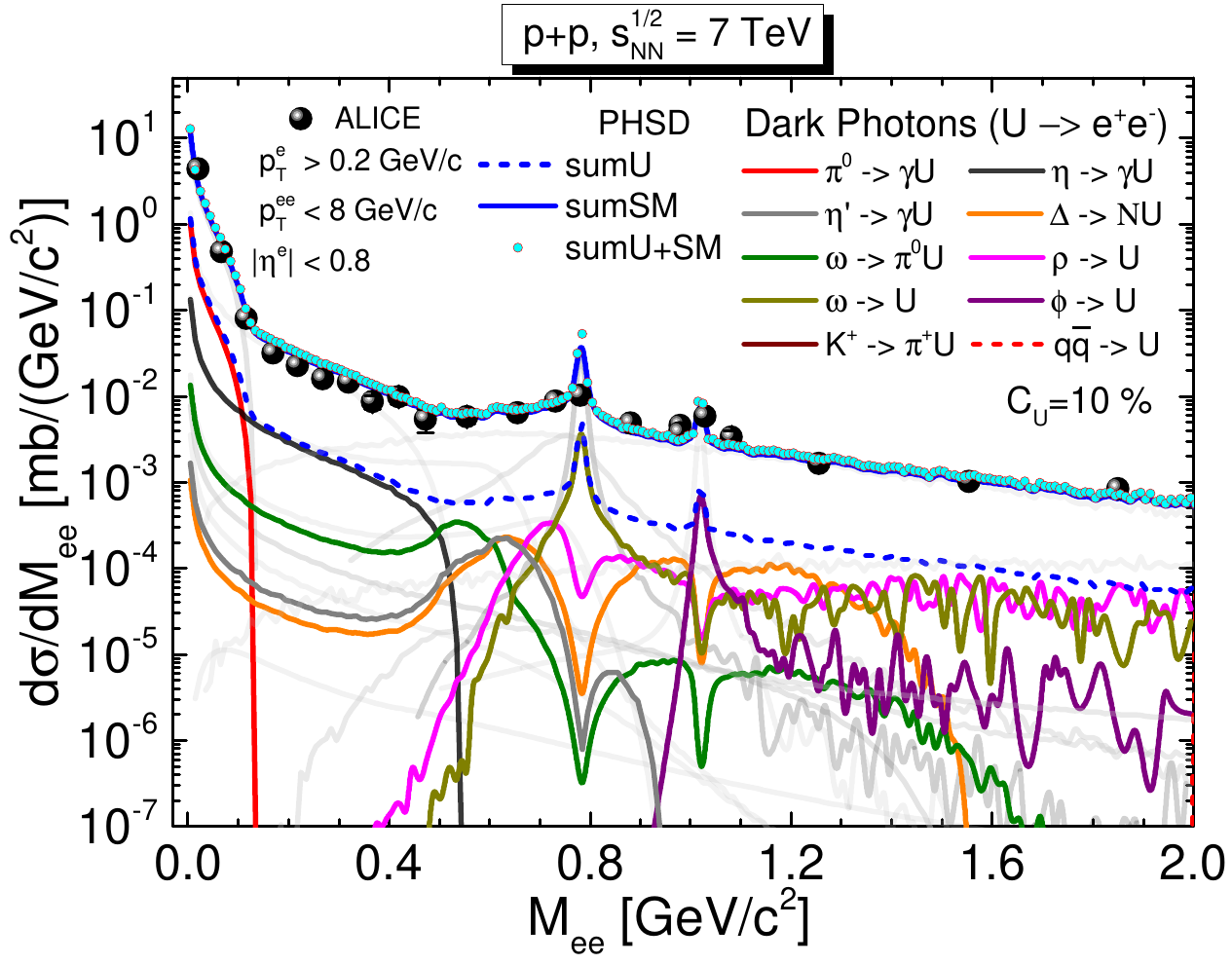}
     \includegraphics[width=0.49\linewidth]{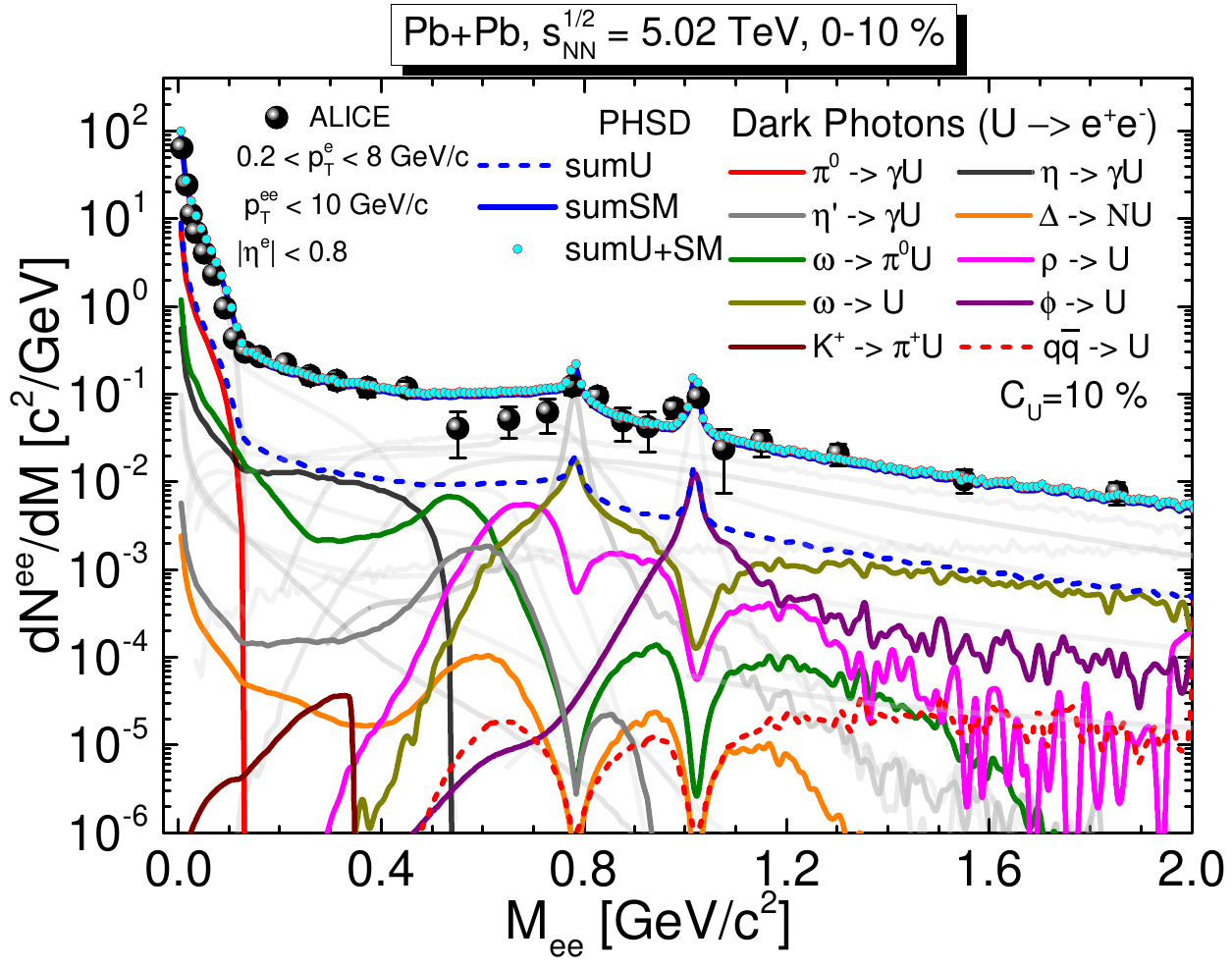}
    \caption{
The invariant mass spectra of dileptons produced in p+p and Pb+Pb collisions at $\sqrt{s_{NN}} = 7$ TeV (left panel) and $5.02$ TeV for 0-10 \% (right panel) respectively, we calculated them using PHSD and compared to ALICE experimental data  ~\cite{ALICE:2018fvj} and   ~\cite{ALICE:2023jef} respectively. The total dilepton yields predicted by the PHSD model are represented by solid blue lines, while the individual contributions from different production channels are detailed in the legends (see legend of Fig. \ref{RHIC_AuAu}).
The ALICE experimental data points are depicted as solid black dots.
To facilitate a direct comparison, the theoretical calculations are adjusted to align with the ALICE acceptance criteria, incorporating its mass and momentum resolution. }
    \label{LHC_pp}
\end{figure*}
Eq. (\ref{epsM}) offers a method to calculate $\varepsilon^2$ for each mass interval $[M_U, M_U + dM]$, which represents the weighted dilepton yield from dark photons in comparison to the contributions from Standard Model processes. 
Thus, the parameter $C_U$ in Eq.~(51) quantifies the allowed contribution of dark photons as a fraction of the total dilepton yield, evaluated bin-by-bin in invariant mass. This condition ensures that the dark photon signal remains within a controlled excess over the SM yield. It should be emphasized, however, that for $\varepsilon^2=1$ the total dark photon spectrum is not identical to the SM spectrum: its shape is determined by the individual decay channels (Eqs.~(\ref{decay1}–\ref{decay10}), weighted by their respective branching ratios. 

This approach also accounts for the experimental acceptance of $e^+e^-$ pairs produced by dark photon decays, treating them similarly to the contributions from Standard Model channels, ensuring compatibility with experimental data. Through a comparison of theoretical predictions with experimental measurements, the possible values for $C_U$, which regulates the extra yield from dark photons, can be determined. Given that dark photons have not been observed in dilepton experiments, any additional yield from these particles must remain within the uncertainties of the experimental data, under the assumption that Standard Model predictions are in agreement with the measurements.

\subsection{Results for the dilepton spectra from dark photon decays }

Fig. \ref{yieldAu200} displays  the number of dark photons
produced per invariant‐mass interval in minimum‐bias Au+Au collisions at
$\sqrt{s_{_{NN}}}=200\,$GeV. A pronounced peak at $M_U\approx m_\omega$
originates from the $\omega\to U$ channel, and another at
$M_U\approx m_\phi$ corresponds to $\phi\to U$. At low masses
($M_U\lesssim0.15\,$GeV/$c^2$), the $\pi^0\to\gamma U$ decay dominates,
producing the initial rise, while the $\eta\to\gamma U$ contribution extends
up to about $0.55\,$GeV/$c^2$. The continuum from $q\bar q\to U$ annihilation
(dashed red) is visible across the mass range but at lower yield compared to
the vector‐meson resonances. The blue dashed line shows the total yield,
demonstrating that meson‐decay processes dominate the $e^+e^-$ emission via
a dark photon for $\varepsilon^2=10^{-6}$.

In Fig.~\ref{M_Hades} the PHSD predictions for the invariant mass spectra of dileptons are shown for four different nuclear systems. The top left and top right panels display the differential cross sections $d\sigma/dM$ for electron-positron pair production in $p+p$ and $p+Nb$ collisions, respectively, both at a beam energy of 3.5 AGeV. The bottom panels illustrate the mass spectra $dN/dM$, normalized to the $\pi^0$ yield, for $Ar+KCl$ collisions at 1.76 AGeV (bottom left) and $Au+Au$ collisions at 1.23 AGeV (bottom right). These theoretical predictions are compared directly with HADES experimental data. In these plots, the individual Standard Model  channels are represented as gray lines, while the total SM contribution is shown as a solid blue line, labeled “Sum SM” in the legend.
The PHSD simulations demonstrate a very good agreement with the HADES measurements across all collision systems, thereby validating and reinforcing the findings reported in Ref. \cite{Bratkovskaya:2013vx,Song:2018xca,Jorge:2025wwp} using the same theoretical framework.

\begin{figure*}[!]
  \centering
\includegraphics[width=0.49\textwidth]{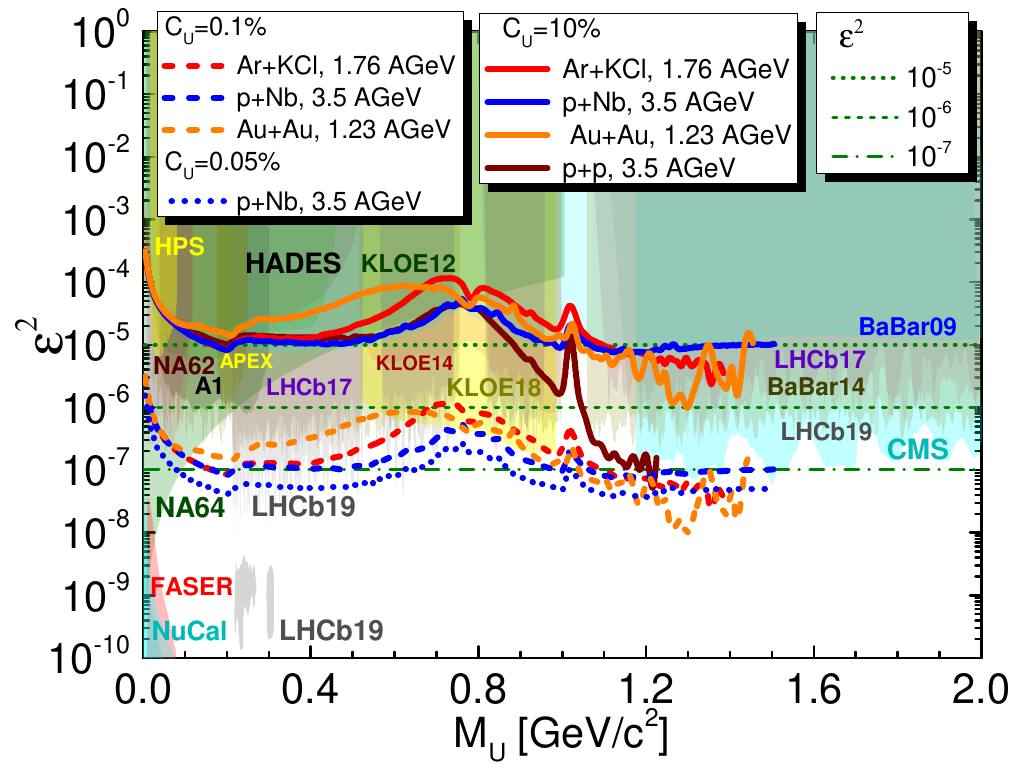}
\includegraphics[width=0.49\textwidth]{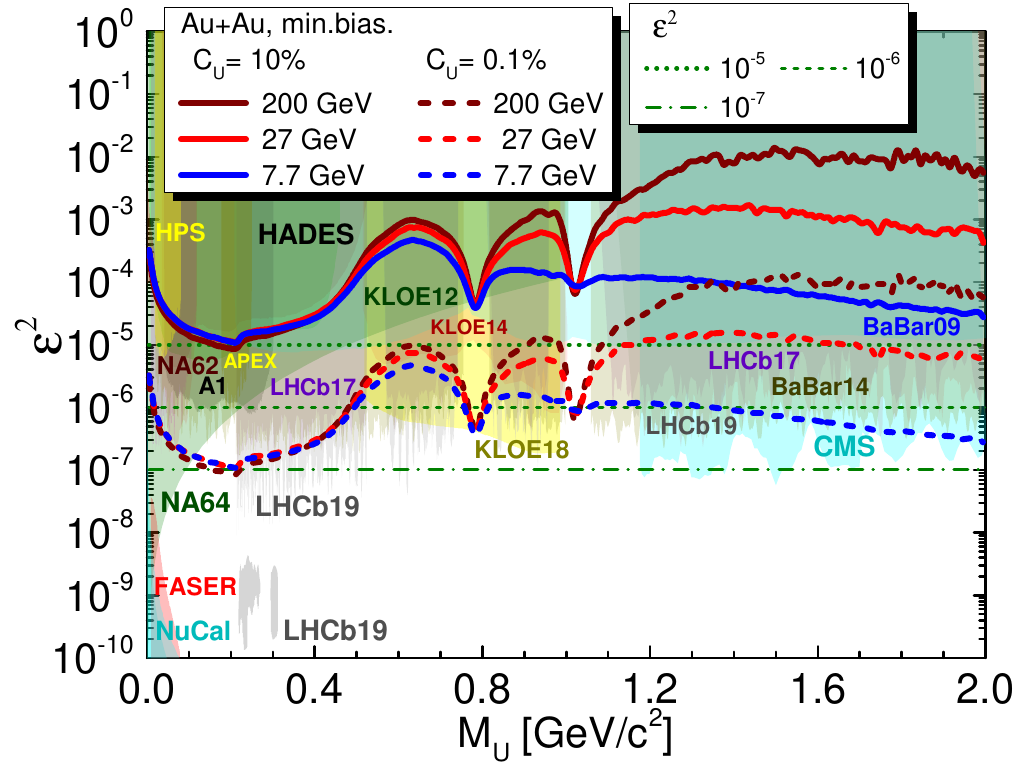}
\includegraphics[width=0.49\textwidth]{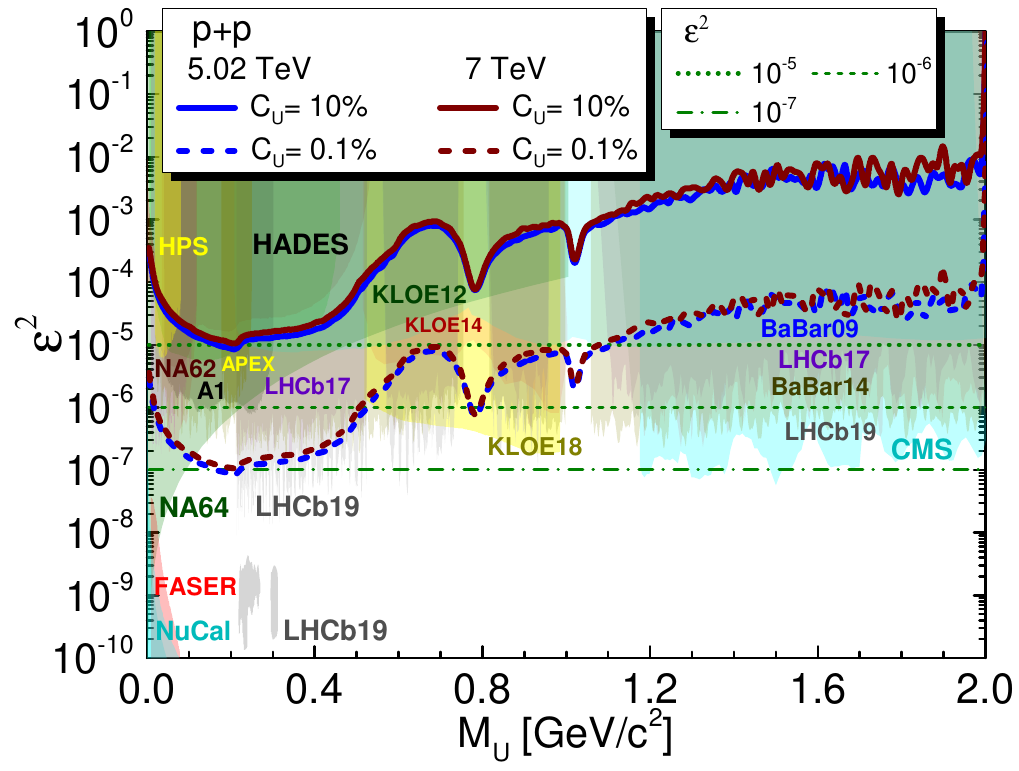}
\includegraphics[width=0.49\textwidth]{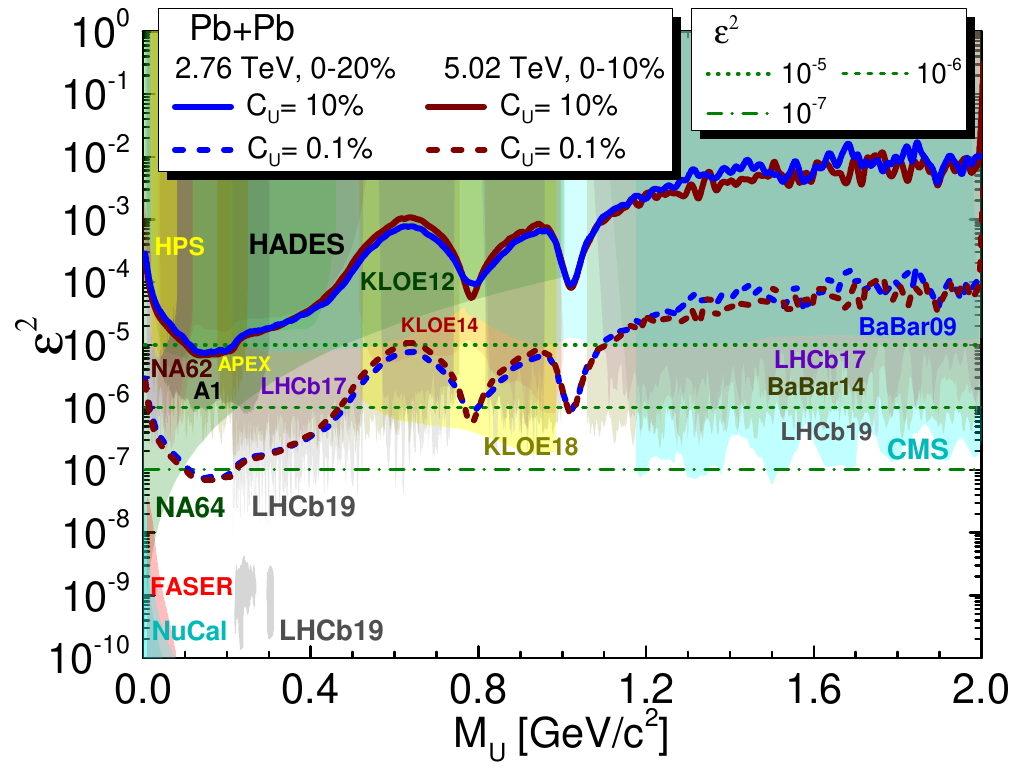}
\caption{
    Kinetic mixing parameter $\epsilon^2 (M_U)$ extracted from PHSD dilepton spectra for various systems and energies:
    (top left panel) at SIS energies: $p+p$ at 3.5 AGeV (brown), $Au+Au$ at 1.23 AGeV (orange), $p+Nb$ at 3.5 AGeV (blue) and $Ar+KCl$ at 1.76 AGeV (red);  (top right panel) at RHIC energies: $Au+Au$ at $\sqrt{s_{NN}}$=7.7 (orange), 27 (blue) and 200 GeV (brown); at LHC energies: (bottom left panel) p+p collisions at  $\sqrt{s_{NN}}$=5.02 (blue) and 7 TeV (brown);   (bottom right panel)  $Pb+Pb$ collisions at $\sqrt{s_{NN}}$=2.76 (blue) and 5.02 TeV (brown) for 0-20\% and 0-10\% centrality respectively.
    In all panels the PHSD lines are shown for a surplus \(C_U=10 \%\) (solid), 0.1\% (dashed) and 0.05\% (dotted).  
    The constant  \(\epsilon^2=10^{-5}\) is shown by dotted olive lines and for \(10^{-6}\) by dashed olive lines for benchmark.  
    The world experimental data for the existing exclusion limits for $\epsilon^2 (M_U)$ are shown for 90\% CL from HADES \cite{Agakishiev:2013fwl}, HPS \cite{HPS:2018xkw}, NA62 \cite{NA62:2019meo}, NA64 \cite{NA64:2019auh}, FASER \cite{FASER:2023tle}, NuCal \cite{Blumlein:2013cua}
    APEX \cite{APEX:2011dww}, A1 at MAMI \cite{Merkel:2014avp}, KLOE \cite{KLOE-2:2016ydq,KLOE-2:2014qxg,KLOE-2:2018kqf}, BaBar \cite{BaBar:2009lbr,BaBar:2014zli}, LHCb \cite{Ilten:2016tkc,LHCb:2017trq,Aaij:2019bvg}  and CMS \cite{CMS:2023hwl}. 
  }
  \label{epsil2}
\end{figure*}

Dilepton contributions from hypothetical dark photon decays are also included in Fig.~\ref{M_Hades}, each represented by different markers. The decay chain $U \rightarrow e^+e^-$ is considered following Dalitz decays of mesons: $\pi^0$ (red), $\eta$ (black), and $\Delta$ resonances (orange), as well as $\omega$ (olive). Direct decays of vector mesons such as $\rho$, $\omega$, and $\phi$ are also incorporated (magenta, dark yellow, and purple, respectively). Their total contribution per mass bin $M = M_U$ is indicated by a dashed blue line labeled "Sum U". The sum of both the SM and dark photon components are marked with cyan dots under the label "Sum SM+U".

The strength of the dark photon contribution is adjusted by a scaling factor $C_U = 0.1$, meaning that up to 10\% more dileptons than expected from SM sources are attributed to dark photon decays in each mass bin. This parameter effectively defines the allowed deviation from the SM prediction due to new physics.

To investigate potential dark photon production at invariant masses above 1 GeV$/c^2$, the analysis is extended to $Au+Au$ collisions at RHIC energies, as well as $p+p$ and $Pb+Pb$ collisions at LHC energies. 
In Fig. \ref{RHIC_AuAu}, the dilepton spectra $dN/dM$ for $\sqrt{s_{NN}} = 27$ GeV (left) and 200 GeV (right) are presented and compared with STAR experimental data.
In Fig. \ref{LHC_pp}, the differential cross section $d\sigma/dM_{ee}$ for $p+p$ at $\sqrt{s_{NN}} = 7$ TeV (left) and the dilepton spectra $dN/dM$ for $Pb+Pb$ at $\sqrt{s_{NN}} = 5.02$ TeV (right)  are displayed and compared with ALICE experimental data.
In both figures, light gray lines represent SM yield background, and the same set of dark photon channels as in Fig. \ref{M_Hades} is included. Notably, the decays $\eta^{\prime} \to \gamma + U$ (gray), $K^+ \to \pi^+ U$ (dark yellow), and $q\bar{q}  \to U$ (dashed red) are also considered at RHIC and LHC energies. 

The $U \rightarrow e^+e^-$ channels are incorporated using the same 10\% surplus assumption. The resulting dark photon signal is shown as a dashed blue line, and the total yield including both SM and dark photon sources is plotted as cyan dots. Experimental data points (STAR and LHC) are shown in black. The theoretical spectra are folded with the experimental acceptance and resolution to enable direct comparison.

As depicted in Fig.~\ref{RHIC_AuAu} and Fig.~\ref{LHC_pp}, the PHSD model, including both SM and dark photon sources, is able to reproduce the experimental spectra over the entire mass range. The PHSD simulations demonstrate a very good agreement with the STAR and ALICE measurements across all collision systems, thus corroborating and supporting the results reported in Ref. \cite{Bratkovskaya:2013vx,Song:2018xca,Jorge:2025wwp} using the same theoretical framework. For invariant masses beyond 1.2 GeV$/c^2$, the primary sources of dark photons are decays of high-mass vector mesons like $\rho$, $\phi$, and $\omega$, along with $\Delta$ resonances.

A notable feature evident in  Figs. \ref{M_Hades}, \ref{RHIC_AuAu} and \ref{LHC_pp} is that the dark photon contributions exhibit the same spectral shape as those of the Standard Model.  
This behavior emerges from the definition of the kinetic mixing parameter $\varepsilon^2(M_U)$ through Eq.(\ref{epsM}), which scales the dark photon contribution by the factor $C_U$ in each mass bin $M_U$, see Eq. (\ref{epsM}).

The role of vector mesons becomes more prominent with increasing collision energy, particularly for masses above 0.6 GeV$/c^2$. Their conversion into dark photons is essential for setting reliable bounds on the kinetic mixing parameter, as discussed in Refs. \cite{Batell:2009yf, Batell:2009di}.

Fig.~\ref{epsil2} presents the upper limits on the kinetic–mixing parameter $\varepsilon^2$ as a function of the dark photon mass $M_U$, extracted from PHSD dilepton spectra across a variety of collision systems and energies.  
The top-left panel shows $\varepsilon^2(M_U)$  at SIS energies: $Ar+KCl$ at 1.76 AGeV (red), $p+p$ at 3.5 AGeV (brown), $p+Nb$ at 3.5 AGeV (blue) and $Au+Au$ at 1.23 AGeV (orange).  
The top right panel shows $\varepsilon^2(M_U)$  at RHIC energies:  Au+Au at $\sqrt{s_{NN}}=7.7$ GeV (blue), 27 GeV (red), and 200 GeV (brown), with the  limitation obtained using STAR data.  
The bottom  row compares $\varepsilon^2(M_U)$ at LHC energies, i.e.  
the bottom left panel: $p+p$ collisions at $\sqrt{s}=5.02$ TeV (blue) and 7 TeV (brown);  
 the bottom right panel:  Pb+Pb at $\sqrt{s_{NN}}=5.02$ TeV (brown) and 2.76 TeV (blue) in 0–10\% and 0–20\% centrality classes, respectively.

In all panels, the PHSD lines correspond to a 10\% allowed dark photon surplus ($C_U=10^{-1}$), while dashed lines denote a 0.1\% surplus ($C_U=10^{-3}$) and dotted lines represent a 0.05\% surplus ($C_U=5 \times 10^{-4}$).  Additionally, the  dotted, dashed and dot-dashed green lines represent constant values of $\varepsilon^2 = 10^{-5}$, $\varepsilon^2 = 10^{-6}$ and $\varepsilon^2 = 10^{-7}$, respectively. The last value approximately corresponds to the current upper limit established by the compilation of global experimental data \cite{Agrawal:2021dbo,Ilten:2018crw}.

The PHSD–derived exclusion limits in Fig.~\ref{epsil2} correspond to a dark photon surplus factor of $C_U=0.10$ (solid lines), with a secondary set at $C_U=0.001$ (dashed lines) for comparison.  This choice ensures that the derived bounds on $\varepsilon^2$ remain robust across the entire mass range, since both Standard Model and dark photon contributions are subject to identical acceptance effects and thus yield invariant line shapes.  

Below the pion threshold ($M_U < m_{\pi^0}$), dark photon production is dominated by the Dalitz decay $\pi^0\to\gamma\,U$, making the extracted $\varepsilon^2$ nearly independent of collision energy or system size. 
However, we note that, following the latest exclusion limits from NA64, NuCal and FASER (cf. Fig.~\ref{epsil2}), these regions have been excluded, and a dark photon production with the mass of $M_U \geq 0.1$ GeV$/c^2$ is more probable.
Above this threshold, however, additional production channels open vector meson conversions, kaon decays, and resonance Dalitz decays which imprint a characteristic mass dependence on the upper limits through their varying contributions to the total dilepton yield.  

For the SIS energy systems ($p+p$, $Ar+KCl$, $p+Nb$ at a few AGeV), the PHSD limits track the shape of the exclusion limits in very good agreement, reflecting minimal contamination from partonic or heavy–flavor sources. 
The extracted upper limit on $\varepsilon^2(M_U)$ for $p+p$ and $p+Nb$ collisions at 3.5\,AGeV agrees well with the BaBar\,09, KLOE12 and HADES results in the mass range $0.2 < M_U < 1\ \mathrm{GeV}/c^2,$  assuming an enhancement factor of $C_U = 10\%$. However, the experimental sensitivities have been surpassed by more recent measurements from BaBar14, KLOE18, LHCb17 and CMS, which have established significantly more stringent upper limits. To remain consistent with these constraints, the dark photon contribution must be limited to $C_U = 0.1\%$, indicating that any potential signal cannot exceed $0.1\%$ of the Standard Model dilepton yield.

In the high-mass region ($M > 1$ GeV$/c^2$), results at RHIC energies ($Au+Au$ at 19 and 200 GeV) and LHC energies ($p+p$, $Pb+Pb$ at TeV scales) show discrepancies with the expected upper limits of the kinetic mixing  shape. The contributions from correlated charm decays, primary Drell-Yan processes ($q\bar{q} \rightarrow U$) and electromagnetic bremsstrahlung ($e^-Z \rightarrow e^-Z U$)
as proposed in Refs.~\cite{Fabbrichesi:2020wbt,Alexander:2016aln,Berlin:2018pwi}, could influence this region. These mechanisms are not yet included but offer promising directions for future studies.

In contrast, the agreement between predicted kinetic mixing phase space and experimental upper limits  is substantially better at lower energies, such as at SIS, where neither primary Drell-Yann decays nor charm production are relevant. This is reflected in the good consistency observed for $p+p$, $Ar+KCl$, and $p+Nb$ systems (see top left panel of Fig. \ref{epsil2}).

However, advances in experimental sensitivity have steadily tightened the upper limit on the kinetic mixing parameter $\varepsilon^2(M_U)$, with the BaBar14, LHCb17, KLOE18 and CMS search providing one of the most stringent exclusions.  To adjust our PHSD limits for $p+Nb$ at 3.5 AGeV and $Ar+KCl$ at 1.76 AGeV into full agreement with the LHCb17, KLOE18 and CMS constraints, the dark photon surplus must be reduced to  $ C_U = 0.001\ (\text{0.1\%}).$
This adjustment produces an approximately constant exclusion line  $\varepsilon^2 \simeq \times10^{-7}$ over the mass range $0.1 < M_U < 1.2\ \mathrm{GeV}/c^2$, matching the experimental upper bound. But following the latest measurements from LHCb19 we require a lower surplus of $C_U = 5 \times 10^{-4}$ ($0.05\%$) in the mass range $0.1 < M_U < 1.2\ \mathrm{GeV}/c^2$, to be below the  experimental upper bound.

\section{Summary}\label{sec:summary}
This work presents a microscopic transport study employing the PHSD approach to evaluate the dilepton yields originating from hypothetical dark photons (or $U$-bosons) decaying into $e^+e^-$ pairs in $p+p$ and $A+A$ collisions from SIS to LHC energies. It builds upon the previous analysis in Refs.~\cite{Schmidt:2021hhs,Bratkovskaya:2022cch}, which focused on dark photon production through $\pi^0 \to \gamma + U$, $\eta \to \gamma + U$, and $\Delta \to N + U$ channels, by incorporating additional processes, such as  the direct decay of vector mesons ($V \to U$, where $V = \rho, \phi, \omega$), the $\omega$ Dalitz decay ($\omega \to \pi^0 + U$), the $\eta^{\prime}$ Dalitz decay ($\eta^{\prime} \to \gamma + U$), the kaon decay ($K^+ \to \pi^+ + U$) and $q\bar q$ annihilation to a dark photon
($q\bar{q} \to U$).

Whenever the dark photon mass lies below the neutral-pion threshold, \(M_U < m_{\pi^{0}}\), production in hadronic reactions is governed almost exclusively by the Dalitz transition \(\pi^{0}\!\to\!\gamma U\). Under these circumstances the kinetic mixing parameter \(\varepsilon^{2}\) deduced from dilepton spectra is essentially insensitive to the beam energy and to the size of the colliding system.  
At $M_U > m_\pi$ further production channels of dark photons become kinematically accessible such as direct conversions of light vector mesons (\(\rho,\omega,\phi\)), strangeness-driven channels such as \(K^{+}\!\to\!\pi^{+}U\), and Dalitz decays of heavier resonances.  The changing balance among these mechanisms with increasing mass imprints a characteristic \(M_U\) dependent pattern on the upper limits extracted for \(\varepsilon^{2}\) through their cumulative contribution to the dilepton spectra.

To establish theoretical upper limits on the kinetic mixing parameter $\varepsilon^2(M_U)$, we employ the procedure developed in Refs. \cite{Schmidt:2021hhs,Bratkovskaya:2022cch}. In the absence of any observed excess attributable to dark photons in current dilepton spectra, the approach imposes an upper bound by requiring that the predicted dark photon contribution does not exceed a predefined surplus level $C_U$, beyond which the dark photons would be distinguishable in the experimental data.

At SIS energies, our results for the kinetic mixing parameter $\varepsilon^2(M_U)$ exhibit close agreement with the BaBar09,  KLOE12 and HADES exclusion limits in the mass range $0.2 < M_U < 1.5$ GeV$/c^2$, assuming a surplus threshold of $C_U = 10\%$. However, more recent experimental data impose tighter constraints, requiring correspondingly lower values of $\varepsilon^2$.
For instance, compatibility with the LHCb17, KLOE18 and CMS upper limits is achieved in the broader mass range $0.2 < M_U < 1.5$ GeV$/c^2$  when the dark photon contribution is limited to $C_U = 0.1\%$, indicating that any potential signal must constitute no more than $0.1\%$ of the Standard Model dilepton yield.
Furthermore, to remain consistent with the latest LHCb measurements, a significantly more stringent constraint of $C_U = 5 \times 10^{-4}$ (i.e., $0.05\%$)  must be imposed in the mass range $0.1 < M_U < 1.2$ GeV$/c^2$.
These results are in agreement with the global set of experimental upper bounds on $\varepsilon^2(M_U)$, further supporting the sensitivity of dilepton measurements in heavy-ion collisions as a probe of dark photon scenarios.

\section*{Acknowledgements}
A.R.J. expresses gratitude for the financial support from the Stiftung Giersch. We also acknowledge the support by the Deutsche Forschungsgemeinschaft (DFG) through the grant CRC-TR 211 "Strong-interaction matter under extreme conditions" (Project number 315477589 - TRR 211) and the CNRS Helmholtz Dark Matter Lab (DMLab). The computational resources utilized for this work were provided by the Center for Scientific Computing (CSC) at Goethe University Frankfurt.

\bibliography{references}

\end{document}